\documentclass[journal,onecolumn]{IEEEtran}
\usepackage{amssymb}
\usepackage{graphicx}
\usepackage{caption}
\usepackage{subcaption}
\usepackage{epstopdf}
\usepackage{amsmath}
\usepackage{amsfonts}
\usepackage{mathrsfs}
\usepackage{cite,url}
\usepackage{units}
\newtheorem{thm}{Theorem}
\newtheorem{theorem}[thm]{Theorem}
\newtheorem{lemma}[thm]{Lemma}
\newtheorem{definition}[thm]{Definition}
\newtheorem{proposition}[thm]{Proposition}
\newtheorem{corollary}[thm]{Corollary}

\newtheorem{example}{Example}

\newtheorem{remark}{Remark}

\newcommand{\I}{\mathscr{I}}
\newcommand{\X}{\mathbb{X}}
\begin{document}
\title {Minimization Problems Based on\\ Relative $\alpha$-Entropy I: Forward Projection}
\author{
M.~Ashok~Kumar and Rajesh~Sundaresan
\thanks{M.~Ashok~Kumar was supported by a Council for Scientific and Industrial Research (CSIR) fellowship and by the Department of Science and Technology. R.~Sundaresan was supported in part by the University Grants Commission by Grant Part (2B) UGC-CAS-(Ph.IV) and in part by the Department of Science and Technology. A part of the material in this paper was presented at the IEEE International Symposium on Information Theory (ISIT 2011), St. Petersburg, Russia, August 2011 \cite{201107ISIT_KumSun}.  A part of the material in the Introduction has overlap with a conference article \cite{201410NCC_KumSun} presented at the National Conference on Communication (NCC 2015), Mumbai, India, held during February 2015.}
\thanks{M.~Ashok~Kumar and R.~Sundaresan are with the ECE Department, Indian Institute of Science, Bangalore 560012, India.}
}

\maketitle

\begin{abstract}
Minimization problems with respect to a one-parameter family of generalized relative entropies are studied. These relative entropies, which we term relative $\alpha$-entropies (denoted $\mathscr{I}_{\alpha}$), arise as redundancies under mismatched compression when cumulants of compressed lengths are considered instead of expected compressed lengths. These parametric relative entropies are a generalization of the usual relative entropy (Kullback-Leibler divergence). Just like relative entropy, these relative $\alpha$-entropies behave like squared Euclidean distance and satisfy the Pythagorean property. Minimizers of these relative $\alpha$-entropies on closed and convex sets are shown to exist. Such minimizations generalize the maximum R\'{e}nyi or Tsallis entropy principle. The minimizing probability distribution (termed forward $\mathscr{I}_{\alpha}$-projection) for a linear family is shown to obey a power-law. Other results in connection with statistical inference, namely subspace transitivity and iterated projections, are also established. In a companion paper, a related minimization problem of interest in robust statistics that leads to a reverse $\mathscr{I}_{\alpha}$-projection is studied.
\end{abstract}


\section{Introduction}
\label{p1:sec:introduction}

Relative entropy\footnote{The relative entropy of $P$ with respect to $Q$ is defined as
\[
  \I(P \| Q) := \sum\limits_{x \in \X} P(x) \log \frac{P(x)}{Q(x)}
\]
and the Shannon entropy of $P$ is defined as
\[
 H(P) := - \sum\limits_{x \in \X} P(x) \log P(x).
\]
The usual convention is $p \log \frac{p}{q} = 0$ if $p = 0$ and $+\infty$ if $p > q = 0$.} or Kullback-Leibler divergence $\mathscr{I}(P\|Q)$ between two probability measures is a fundamental quantity that arises in a variety of situations in probability theory, statistics, and information theory. In probability theory, it arises as the rate function for estimating the probability of a large deviation for the empirical measure of independent samplings. In statistics, for example, it arises as the best error exponent in deciding between two hypothetical distributions for observed data. In Shannon theory, it is the penalty in expected compressed length, namely the gap from Shannon entropy $H(P)$, when the compressor assumes (for a finite-alphabet source) a mismatched probability measure $Q$ instead of the true probability measure $P$.

Relative entropy also brings statistics and probability theory together to provide a foundation for the well-known maximum entropy principle for decision making under uncertainty. This is an idea that goes back to L.~Boltzmann, was popularized by E.~T.~Jaynes \cite{1982xxPPSSP_Jay}, and has its foundation in the theory of large deviation. Suppose that an ensemble average measurement (say sample mean, sample second moment, or any other similar linear statistic) is made on the realization of a sequence of independent and identically distributed (i.i.d.) random variables. The realization must then have an empirical measure that obeys the constraint placed by the measurement -- the empirical measure must belong to an appropriate convex set, say $\mathbb{E}$. Large deviation theory tells us that a special member of $\mathbb{E}$, denoted $P^*$, is overwhelmingly more likely than the others. If the alphabet $\mathbb{X}$ is finite (with cardinality $|\mathbb{X}|$), and the prior probability (before measurement) is the uniform measure $U$ on $\mathbb{X}$, then $P^*$ is the one that minimizes the relative entropy
\[
  \mathscr{I}(P \| U) = \log |\mathbb{X}| - H(P),
\]
which is the same as the one that maximizes (Shannon) entropy, subject to $P \in \mathbb{E}$. This explains why the principle is called maximum entropy principle. In Jaynes' words, \emph{``... it is maximally noncommittal to the missing information''} \cite{1982xxPPSSP_Jay}.

As a physical example, let us tag a particular molecule in the atmosphere. Let $X$ denote the height of the molecule in the atmosphere. Then the potential energy of the molecule is $mgX$. Let us suppose that the average potential energy is held constant, that is, $E[mgX] = c$, a constant. Then the probability distribution of the height of the molecule is taken to be the exponential distribution $\lambda \exp{(-\lambda x)}$, where $\lambda = mg/c$. This is also the maximum entropy probability distribution subject to first moment constraint \cite{2006xxEIT_CovTho}. 

More generally, if the prior probability (before measurement) is $Q$, then $P^*$ minimizes $\mathscr{I}(P \| Q)$ subject to $P \in \mathbb{E}$. Something more specific can be said: $P^*$ is the limiting conditional distribution of a ``tagged'' particle under the conditioning imposed by the measurement. This is called \emph{the conditional limit theorem} or the {\em Gibbs conditioning principle}; see for example Campenhout and Cover \cite{1981TIT_Cam_Cov} or Csisz\'ar \cite{1984xxAP_Csi} for a more general result.

It is well-known that $\mathscr{I}(P\|Q)$ behaves like ``squared Euclidean distance'' and has the ``Pythagorean property'' (Csisz\'ar \cite{1975xxAP_Csi}). In view of this and since $P^*$ minimizes $\mathscr{I}(P \| Q)$ subject to $P \in \mathbb{E}$, one says that $P^*$ is ``closest'' to $Q$ in the relative entropy sense amongst the measures in $\mathbb{E}$, or in other words, ``$P^*$ is the forward $\mathscr{I}$-projection of $Q$ on $\mathbb{E}$''. Motivated by the above maximum entropy and Gibbs conditioning principles, $\mathscr{I}$-projection was extensively studied by Csisz\'{a}r \cite{1984xxAP_Csi}, \cite{1975xxAP_Csi}, Csisz\'{a}r and Mat\'{u}\v{s} \cite{200306TIT_CsiMat}, Csisz\'ar and Shields \cite{2004xxITST_CsiShi}, and Csisz\'{a}r and Tusn\'{a}dy \cite{1984SAD_CsiTus}. More recently, minimizations of general entropy functionals with convex integrands were studied by Csisz\'{a}r and Mat\'{u}\v{s} \cite{2012xxKYB_CsiMat}. These include Bregman's divergences and Csisz\'ar's $f$-divergences. $\mathscr{I}$-minimization also arises in the contraction principle in large deviation theory (see for example Dembo and Zeitouni's \cite[p.126]{1998xxLDTA_DemZei}).

This paper is on projections or minimization problems associated with a parametric generalization of relative entropy. To see how this parametric generalization arises, we return to our remark on how relative entropy arises in Shannon theory. For this, we must first recall how R\'enyi entropies are a parametric generalization of the Shannon entropy.

R\'{e}nyi entropies $H_{\alpha}(P)$ for $\alpha \in (0, 1)$ play the role of Shannon entropy when the \emph{normalized cumulant} of compression length is considered instead of expected compression length. Campbell \cite{1965xxIC_Cam} showed that
\[
  \min \frac{1}{n\rho} \log E \left[ \exp \{ \rho L_n (X^n) \} \right] \to H_{\alpha} (P) ~ (\mbox{as } n \to \infty)
\]
for an i.i.d. source with marginal $P$. The minimum is over all compression strategies $L_n$ that satisfy the Kraft inequality\footnote{A compression strategy $L_n\colon \X^n\to \{0,1,2,\dots\}$ assigns a target codeword length $L_n(x^n)$ to each string $x^n\in \X^n$.}, $\alpha = 1/(1+\rho)$, and $\rho > 0$ is the cumulant parameter. We also have $\lim_{\alpha \to 1} H_{\alpha}(P) = H(P)$, so that R\'{e}nyi entropy may be viewed as a generalization of Shannon entropy.

If the compressor assumed that the true probability measure is $Q$, instead of $P$, then the gap in the normalized cumulant's optimal value is an analogous parametric divergence quantity\footnote{Blumer and McEliece \cite{198809TIT_BluMcE}, in their attempt to find better upper and lower bounds on the redundancy of generalized Huffman coding, were indirectly bounding this parameterized divergence.}, which we shall denote $\mathscr{I}_{\alpha}(P,Q)$ \cite{200701TIT_Sun}. The same quantity\footnote{We suggest the pronunciation ``I-alpha'' for $\I_{\alpha}$.} also arises when we study the gap from optimality of mismatched guessing exponents. See Arikan \cite{199601TIT_Ari} and Hanawal and Sundaresan \cite{201101TIT_HanSun} for general results on guessing, and see Sundaresan \cite{200206ISIT_Sun},\cite{200701TIT_Sun} on how $\mathscr{I}_{\alpha}(P,Q)$ arises in the context of mismatched guessing. Recently, Bunte and Lapidoth \cite{2014xxarx_BunLap} have shown that the $\mathscr{I}_{\alpha}(P,Q)$ also arises as redundancy in a mismatched version of the problem of coding for tasks.

As one might expect, it is known that (see for example, Sundaresan \cite[Sec.~V-5)]{200701TIT_Sun} or Johnson and Vignat \cite[A.1]{200705AIHP_JohVig}) $\lim_{\alpha \to 1} \mathscr{I}_{\alpha}(P,Q) = \mathscr{I}(P\|Q)$, so that we may think of relative entropy as $\mathscr{I}_1(P,Q)$. Thus $\mathscr{I}_{\alpha}$ is a generalization of relative entropy, i.e., a \emph{relative $\alpha$-entropy}\footnote{This terminology is from Lutwak, et al. \cite{200501TIT_LutYanZha}.}.

Not surprisingly, the maximum R\'{e}nyi entropy principle has been considered as a natural alternative to the maximum entropy principle of decision making under uncertainty. This principle is equivalent to another principle of maximizing the so-called Tsallis entropy which happens to be a monotone function of the R\'{e}nyi entropy. R\'{e}nyi entropy maximizers under moment constraints are distributions with a power-law decay (when $\alpha < 1$). See Costa et al. \cite{200307EMMCVPR_CosHerVig} or Johnson and Vignat \cite{200705AIHP_JohVig}. Many statistical physicists have studied this principle in the hope that it may ``explain'' the emergence of power-laws in many naturally occurring physical and socio-economic systems, beginning with Tsallis \cite{1988xxJSP_Tsa}. Based on our explorations of the vast literature on this topic, we feel that our understanding, particularly one that ought to involve a modeling of the \emph{dynamics} of such systems with the observed power-law profiles as equilibria in the asymptotics of large time, is not yet as mature as our understanding of the classical Boltzmann-Gibbs setting. But, by noting that $\mathscr{I}_{\alpha} (P,U) = \log |\mathbb{X}| - H_{\alpha}(P)$, we see that both the maximum R\'{e}nyi entropy principle and the maximum Tsallis entropy principle are particular instances of a ``minimum relative $\alpha$-entropy principle'': 
\[
 \text{ minimize } \mathscr{I}_{\alpha}(P,Q) \text{ over } P \in \mathbb{E}.
\]
We shall call the minimizing $P^*$ as the forward $\mathscr{I}_{\alpha}$-projection of $Q$ on $\mathbb{E}$.

The main aim of this paper is to study forward $\mathscr{I}_{\alpha}$-projections in general measure spaces. Our main contributions are on existence, uniqueness, and structure of these projections. We have several motivations to publish our work.
\begin{itemize}
  \item We provide a rather general sufficient condition on the constraint set under which a forward $\mathscr{I}_{\alpha}$-projection exists and is unique. This can enable statistical physicists to speak of the R\'enyi entropy maximizer and explore its properties even if the maximizer is not known explicitly. While the existence and uniqueness of $\mathscr{I}_{\alpha}$-projection for closed convex sets $\mathbb{E}$ was shown for the finite alphabet case by Sundaresan \cite{200701TIT_Sun}, here we study more general measure spaces (for example $\mathbb{R}^n$).

  \item Unlike relative entropy, its generalization relative $\alpha$-entropy does not, in general, satisfy the well-known data processing inequality, nor is it in general convex in either of its arguments. Nevertheless, there is a remarkable parallelism between relative entropy and relative $\alpha$-entropy. In particular, they share the ``Pythagorean property'' and behave like squared Euclidean distance. This too was explored by Sundaresan \cite{200701TIT_Sun} for the finite alphabet case, and we wish to extend the parallels to more general alphabet spaces.

  \item We provide information on the structure of the R\'enyi entropy maximizer, under linear statistical constraints, whenever the maximizer exists. This can provide statistical physicists a quick means to check if their empirical observations in a particular physical setting conform to the maximum R\'enyi entropy principle. It also provides a means to estimate the appropriate $\alpha$ for a particular physical setting. Interestingly, the R\'enyi entropy maximizers belong to a ``power-law family'' of distributions that are the natural parametric generalizations of the Shannon entropy maximizers, namely the exponential family of distributions.

  \item In a companion paper, we shall show that a robust parameter estimation problem is a ``reverse $\mathscr{I}_{\alpha}$-projection'' problem, where the minimization is with respect to the \emph{second} argument of $\mathscr{I}_{\alpha}$. If this reverse projection is on a power-law family, then one may turn the reverse projection into a forward projection of a specific distribution on an appropriate linear family. In that paper we shall also explore the geometric relationship between the power-law and the linear families.

  \item One may think of the maximum entropy principle or the minimization of relative entropy as a ``projection rule''; see Section \ref{p1:sec:transitivity} for projection rules with some desired properties. Three of these properties are ``regularity'', ``locality'', and ``subspace-transitivity''. It turns out that the $\mathscr{I}_{\alpha}$-based projection rule is regular, subspace-transitive when $\alpha <1$, but ``nonlocal''. Any regular, subspace-transitive, and local projection rule is generated by Bregman's divergences of the sum-form \cite{1991xxTAS_Csi}. In our, as yet not very successful, attempt to characterize all regular, subspace-transitive, but possibly nonlocal projection rules, we wished to understand as much as we could about a particular nonlocal projection rule. The understanding we have gained may be of use to the wider community interested in axiomatic approaches to abstract inference problems.
\end{itemize}

It is known (see for example \cite{200701TIT_Sun}) that $\mathscr{I}_{\alpha}(P,Q)$ is the more commonly studied R\'{e}nyi divergence of order $1/\alpha$, not of the original measures $P$ and $Q$, but of their escort measures $P'$ and $Q'$, where $P'(x) = P(x)^{\alpha} / Z(P)$, and $Z(P)$ is the normalization that makes $P'$ a probability measure. $Q'$ is similarly defined. While the R\'{e}nyi divergences arise naturally in hypothesis testing problems (see for example Csisz\'{a}r \cite{199501TIT_Csi}), $\mathscr{I}_{\alpha}$ arises more naturally as a redundancy for mismatched compression, as discussed earlier. Moreover, $\mathscr{I}_{\alpha}(P,Q)$ is a certain monotone function of Csisz\'{a}r's $f$-divergence between $P'$ and $Q'$. As a consequence of the appearance of the escort measures, the data-processing property satisfied by the $f$-divergences does not hold for the $\mathscr{I}_{\alpha}$-divergences. It is therefore all the more intriguing that it is neither the $f$-divergences nor the R\'enyi divergences but the $\mathscr{I}_{\alpha}$-divergences that share the Pythagorean property with relative entropy. However, quite recently, van Erven and Harremo\"es \cite{201407TIT_ErvHar} showed that R\'enyi divergences have a Pythagorean property when the forward projection is on a so-called {\em $\alpha$-convex set}.



The paper is organized as follows. In Section \ref{p1:sec:alpha_relative_entropy}, we formally define $\mathscr{I}_{\alpha}$ and establish some of its basic algebraic and topological properties, those desired of an information divergence. In Section \ref{p1:sec:projection}, we establish the existence of $\mathscr{I}_{\alpha}$-projection on closed (in an appropriate topology) and convex sets. The proof for the case $\alpha<1$ is analogous to that for relative entropy \cite[Th.~2.1]{1975xxAP_Csi}. The proof for the case $\alpha>1$ exploits some functional analytic tools. In Section \ref{p1:sec:pythagorenproperty}, we present the Pythagorean property in generality and derive some of its immediate consequences in connection with the forward projection. In Section \ref{p1:sec:projection_for_linear_family}, we characterize the forward $\mathscr{I}_{\alpha}$-projection on a linear family of probability measures, whenever it exists. In Section \ref{p1:sec:transitivity}, we establish a desirable subspace transitivity property and further prove the convergence of an iterative method for finding the forward $\mathscr{I}_{\alpha}$-projection on linear families. In the concluding Section \ref{p1:sec:open-questions}, we highlight some interesting open questions.

The companion paper \cite{2014xxManuscript2_KumSun} will explore the orthogonality between the power-law and the linear families, will exploit this orthogonality in a robust parameter estimation problem, and will study the reverse $\mathscr{I}_{\alpha}$-projection in detail.

\section{The relative $\alpha$- entropy}
\label{p1:sec:alpha_relative_entropy}

We begin by defining relative $\alpha$-entropy on a general measure space for all $\alpha>0$ except $\alpha = 1$. As $\alpha \rightarrow 1$ our definition will approach the usual relative entropy or Kullback-Leibler divergence.

Let $P$ and $Q$ be two probability measures on a measure space $(\mathbb{X},\mathcal{X})$. Let $\alpha \in (0,\infty)$ with $\alpha \neq 1$. Let $\mu$ be a dominating $\sigma$-finite measure on $(\mathbb{X},\mathcal{X})$ with respect to which $P$ and $Q$ are both absolutely continuous, denoted $P \ll \mu$ and $Q \ll \mu$. Write $p = dP/d\mu$ and $q = dQ/d\mu$ and assume that $p$ and $q$ belong to the complete topological vector space $L^{\alpha}(\mu)$ with metric
\begin{eqnarray*}
 d(h,g) =
\begin{cases} \left( \int |h - g|^{\alpha} d\mu \right)^{1/\alpha} & \text{ if } \alpha>1,\\
 \int |h - g|^{\alpha} d\mu & \text{ if } \alpha<1.
\end{cases}
\end{eqnarray*}
We shall use the notation
$$\| h \| := \left( \int |h|^{\alpha} d\mu \right)^{1/\alpha}$$
even though $\|\cdot\|$, as defined, is not a norm for $\alpha < 1$. For convenience we suppress the dependence of $d(\cdot,\cdot)$ and $\|\cdot\|$ on $\alpha$; but this dependence should be borne in mind. Throughout we shall restrict attention to probability measures whose densities with respect to $\mu$ are in $L^{\alpha}(\mu)$. The R\'{e}nyi entropy of $P$ of order $\alpha$ (with respect to $\mu$) is defined to be
\begin{equation}
  \label{p1:eqn:renyi-entropy}
  H_{\alpha}(P) := \frac{1}{1-\alpha} \log \left( \int_{\mathbb{X}} p^\alpha d\mu \right).
\end{equation}
Consider the escort measures $P'$ and $Q'$ having densities $p'$ and $q'$ with respect to $\mu$ defined by
\begin{eqnarray}
\label{p1:eqn:escort}
  \frac{dP'}{d\mu} = p' := \frac{p^{\alpha}}{\int p^{\alpha}d\mu} \mbox{ and } \frac{dQ'}{d\mu} = q' := \frac{q^{\alpha}}{\int q^{\alpha}d\mu}.
\end{eqnarray}
Once again, the dependence of $p'$ and $q'$ on $\alpha$ is suppressed for convenience. By setting $\alpha = \frac{1}{1+\rho}$, we have the re-parametrization in terms of $\rho$ with $-1<\rho<\infty$, $\rho \neq 0$, and $\rho = \alpha^{-1}-1$. Define
\begin{equation*}
\label{p1:eqn:the-function-in_I_f}
 f(u) := \text{sgn}(\rho) \cdot (u^{1+\rho} - 1), \quad u \geq 0.
\end{equation*}
Csisz\'{a}r's $f$-divergence \cite{1967xxSSMH1_Csi} between two probability measures $P$ and $Q$, both absolutely continuous with respect to $\mu$, is given by
\begin{equation}
  \label{p1:eqn:If-defn}
  I_{f}(P,Q) :=  \int q f \left( \frac{p}{q} \right) d\mu.
\end{equation}
In the above definition we use the following conventions:
\begin{eqnarray*}
  0\cdot f\left(\frac{0}{0}\right)=0,	
\end{eqnarray*}
and for $a>0$,
\begin{equation*}
  0\cdot f\left(\frac{a}{0}\right)=
   \begin{cases}
     \infty & \text{if $\rho > 0$,} \\
     0 &\text{if $\rho < 0$.}
   \end{cases}
\end{equation*}
Since $f$ is strictly convex when $\rho \neq 0$, by Jensen's inequality, $I_f(P,Q) \geq 0$ with equality if and only if $P=Q$.
\begin{definition}[Relative $\alpha$-entropy] The \textit{$\alpha$-entropy of $P$ relative to $Q$} (or relative $\alpha$-entropy of $P$ with respect to $Q$, or simply relative $\alpha$-entropy) is defined as
 \begin{eqnarray}
\label{p1:eqn:alphadiv}
  \mathscr{I}_{\alpha}^{\mu}(P,Q) := \frac{1}{\rho} \log\left[ \text{sgn}(\rho) \cdot I_{f}(P',Q') + 1\right].
\end{eqnarray}
\end{definition}
$\mathscr{I}_{\alpha}^{\mu}$ depends on the reference measure $\mu$ because the densities $p'$ and $q'$ defined in (\ref{p1:eqn:escort}) do. However, for brevity, we omit the superscript $\mu$ and ask the reader to bear the dependence on $\mu$ in mind. For the information theoretic and statistical physics motivating examples in Section I, $\mu$ is the counting measure or the Lebesgue measure depending on whether $\X$ is finite or $\mathbb{R}^d$.

From the conventions used to define $I_f$, we have $\mathscr{I}_{\alpha}(P,Q) = \infty$ when either
\begin{itemize}
  \item $\alpha < 1$ and $P \not\ll Q$, or
  \item $\alpha > 1$ and $P$ and $Q$ are mutually singular.
\end{itemize}
Abusing notation a little, when speaking of densities, we shall some times write $\mathscr{I}_{\alpha}(p,q)$ for $\mathscr{I}_{\alpha}(P,Q)$. Let us reemphasize that implicit in our definition of $\mathscr{I}_{\alpha}(P,Q)$ is the assumption that $p$ and $q$ are both in $L^{\alpha}(\mu)$.

The following are some alternative expressions of $\mathscr{I}_{\alpha}$ that are used in this paper:
\begin{eqnarray}
\label{p1:alphadiv_linear_form_general}
\mathscr{I}_{\alpha}(P,Q) & = & \frac{\alpha}{1-\alpha} \log \int \frac{p}{\|p\|} \left( \frac{q}{\|q\|} \right)^{\alpha-1} d\mu\\
\label{p1:alphadiv_expanded_general}
 & = & \frac{\alpha}{1-\alpha} \log \int p q^{\alpha-1}d\mu-\frac{1}{1-\alpha}\log \int p^{\alpha}d\mu + \log \int q^{\alpha}d\mu.
\end{eqnarray}
When $\mathbb{X}$ is discrete (with $\mu$ being the counting measure on $\mathbb{X}$), the probability measures may be viewed as finite or countably infinite dimensional vectors. In this case, we may write
\begin{eqnarray}
\label{p1:alphadiv_linear_form_discrete}
\mathscr{I}_{\alpha}(P,Q) & = & \frac{\alpha}{1-\alpha} \log \left[ \sum_x \frac{P(x)}{\|P\|} \left( \frac{Q(x)}{\|Q\|} \right)^{\alpha-1} \right]\\ \label{p1:alphadiv_expanded_discrete}
& = & \frac{\alpha}{1-\alpha} \log \Big[ \sum_x P(x) Q(x)^{\alpha-1} \Big] - \frac{1}{1-\alpha}\log \sum_x P(x)^{\alpha} + \log \sum_x Q(x)^{\alpha}.
\end{eqnarray}
We now summarize some properties of relative $\alpha$-entropy.
\begin{lemma}
\label{p1:lemma:properties}
The following properties hold.
\begin{itemize}
 \item [a)] ({\em Positivity}). $\mathscr{I}_{\alpha}(P,Q) \geq 0$ with equality if and only if $P = Q$.

\vspace{.1in}

 \item[b)] ({\em Generalization of relative entropy}). Let $\mathscr{I}_{\alpha}(P,Q)<\infty$ for some $\alpha=\alpha_l<1$ and simultaneously for some $\alpha = \alpha_u > 1$. Then $\mathscr{I}_{\alpha}(P,Q)$ is well-defined for all $\alpha \in [\alpha_l,\alpha_u] \setminus \{1\}$, and
     \[
       \lim_{\alpha \to 1} \mathscr{I}_{\alpha}(P,Q) = \mathscr{I}(P \| Q),
     \]
     where $\mathscr{I}(P\|Q)$ is the relative entropy of $P$ with respect to $Q$.

\vspace{.1in}

 \item[c)] ({\em Relation to R\'{e}nyi divergence}).
 \[
   \mathscr{I}_{\alpha}(P,Q) = D_{1/\alpha}(P'\|Q'),
 \]
 where
 \[
   D_{\beta}(P\|Q) := \frac{1}{\beta-1} \log \int p^{\beta} q^{1-\beta}d\mu
 \]
 is the R\'{e}nyi divergence of order $\beta$.

\vspace{.1in}

%
%
%

 \item[d)] ({\em Relation to R\'enyi entropy}). Let $|\mathbb{X}|<\infty$ and let $U$ be the uniform probability measure on $\mathbb{X}$. Then $\mathscr{I}_{\alpha}(P,U) = \log |\mathbb{X}| - H_{\alpha}(P).$

\vspace{.1in}

 \item[e)] ({\em R\'enyi entropy maximizer under a covariance constraint}). Let $\mathbb{X} = \mathbb{R}^n$ and let $\mu$ be the Lebesgue measure on $\mathbb{R}^n$. For $\alpha > n / (n+2)$ and $\alpha \neq 1$, define the constant $b_{\alpha} = (1-\alpha)/(2\alpha - n(1-\alpha))$. With $C$ a positive definite covariance matrix, the function
      \[
        \phi_{\alpha, C}(x) = Z_{\alpha}^{-1} \left[ 1 + b_{\alpha} \cdot x^T C^{-1} x \right]^{\frac{1}{\alpha-1}}_+,
      \]
with $[a]_+ := \max \{a,0 \}$ and $Z_{\alpha}$ the normalization constant, is the density function of a probability measure on $\mathbb{R}^n$ whose covariance matrix is $C$. Furthermore, if $g$ is the density function of any other random vector with covariance matrix $C$, then
 \begin{equation}
   \label{p1:eqn:moment_entropy}
    \mathscr{I}_{\alpha}(g, \phi_{\alpha,C}) = H_{\alpha}(\phi_{\alpha,C}) - H_{\alpha}(g).
  \end{equation}
Consequently $\phi_{\alpha,C}$ is the density function of the R\'{e}nyi entropy maximizer among all $\mathbb{R}^n$-valued random vectors with covariance matrix $C$.

\begin{IEEEproof}
See Appendix \ref{p1:sec:app-lemma1}.
\end{IEEEproof}

\end{itemize}
\end{lemma}

\vspace{.1in}

\begin{remark}
 For relative entropy ($\alpha = 1$), the analog of (\ref{p1:eqn:moment_entropy}) under a covariance constraint is $$\mathscr{I}(g \| \phi) = H(\phi) - H(g),$$ where $H$ is differential entropy and $\phi$ is the Gaussian distribution with the same covariance as $g$ \cite[Th.~8.6.5]{2006xxEIT_CovTho}. In Section \ref{p1:sec:projection_for_linear_family} we shall study R\'enyi entropy maximizers under more general linear constraints.
\end{remark}

\vspace{.1in}

\begin{remark}
 While the numerical value of relative entropy $\mathscr{I}(P \| Q)$ does not depend on the dominating measure $\mu$, recall that $\mathscr{I}_{\alpha}(P, Q)$ does depend on $\mu$ in general.
\end{remark}

\vspace{.1in}

Analogous to the property that $p\mapsto \mathscr{I}(p\|q)$ is lower semicontinuous in the topology on $L^1(\mu)$ arising from the total variation metric \cite[Sec.~2.4, Assertion 5]{1964xxIISRV_Pin}, we have the following.

\vspace{.1in}

\begin{proposition}[Lower semicontinuity in the first argument]
\label{p1:prop:lsc1}
For a fixed $q$, consider $p \mapsto \mathscr{I}_{\alpha}(p,q)$ as a function on $L^{\alpha}(\mu)$. This function is continuous for $\alpha>1$ and lower semicontinuous for $\alpha < 1$.
\end{proposition}

\begin{IEEEproof}
See Appendix \ref{p1:sec:app-prop2}.
\end{IEEEproof}

\vspace{.1in}

\begin{remark}
\label{p1:cont_rem}
When $\alpha < 1$, $\mathscr{I}_{\alpha}(\cdot,Q)$ is lower semicontinuous, but not necessarily continuous. To see this, let $\mathbb{X}$ be finite. Let $P_n, P, Q$ be probability measures on $\mathbb{X}$ such that all $P_n$'s have full support, i.e., $P_n(x)>0$ for all $x \in \mathbb{X}$, but $Q(x_0)=0$ for some $x_0 \in \mathbb{X}$, $P \ll Q$, and finally $P_n\to P$. Then $\mathscr{I}_{\alpha}(P_n,Q)=\infty$ for all $n$, but $\mathscr{I}_{\alpha}(P,Q)<\infty$.
\end{remark}

\vspace*{.1in}

\begin{remark}
If however $\mathbb{X}$ is finite and $Q$ has full support, then $\mathscr{I}_{\alpha}(\cdot, Q)$ is indeed continuous and this can be seen by taking the limit term by term in (\ref{p1:alphadiv_linear_form_discrete}).
\end{remark}

\vspace*{.1in}

We now address the behavior as a function of $q$.

\vspace{.1in}

\begin{proposition}
\label{p1:lsc2}
Fix $\alpha > 0$, $\alpha \neq 1$. For a fixed $p$, the mapping $q \mapsto \mathscr{I}_{\alpha}(p,q)$ is lower semicontinuous in $L^{\alpha}(\mu)$.
\end{proposition}

\begin{IEEEproof}
 See Appendix \ref{p1:sec:app-prop3}
\end{IEEEproof}

\vspace*{.1in}

\begin{remark}
When $\mathbb{X}$ is finite, with $+\infty$ as a potential limiting value, $\mathscr{I}_{\alpha}(P,\cdot)$ is continuous for all $\alpha>0$, $\alpha \neq 1$, as is easily seen by taking term-wise limits in the summation in (\ref{p1:alphadiv_linear_form_discrete}).
\end{remark}

\vspace*{.1in}

We next establish quasi-convexity of $\mathscr{I}_{\alpha}$ in the first argument, i.e., for every fixed $q$ and real number $\tau$, the lower level sets $\overline{B}(q,\tau) := \{p\colon\mathscr{I}_{\alpha}(p,q)\le \tau\}$ (or ``$\I_{\alpha}$-balls'') are convex.

\vspace*{.1in}

\begin{proposition}
\label{p1:quasiconvexity}
Fix $\alpha > 0$, $\alpha \neq 1$. For a fixed $q$, the mapping $p \mapsto \mathscr{I}_{\alpha}(p,q)$ is quasi-convex in $L^{\alpha}(\mu)$.
\end{proposition}

\begin{IEEEproof}
See Appendix \ref{p1:sec:app-prop4}
\end{IEEEproof}

\vspace*{.1in}

\begin{remark}
In general, for both $\alpha <1$ and $\alpha >1$, $\mathscr{I}_{\alpha}$ is not convex in either of its arguments. Moreover, $\mathscr{I}_{\alpha}$ does not satisfy the data processing inequality while relative entropy and more generally Csisz\'{a}r's $f$-divergences do.
\end{remark}

\section{Existence and Uniqueness of the Forward $\mathscr{I}_{\alpha}$-projection}
\label{p1:sec:projection}

In this section, we shall introduce the notion of a forward $\mathscr{I}_{\alpha}$-projection of a probability measure on a subset of probability measures. We shall also prove a sufficiency result for the existence of the forward $\mathscr{I}_{\alpha}$-projection. We begin by first proving a useful inequality relating $f$-divergences. This is an inequality that turns out to be the analog of the parallelogram identity of \cite{1975xxAP_Csi} for relative entropy ($\alpha = 1$) and the analog of the Apollonius Theorem in plane geometry (see, for e.g., Bhatia \cite[p.~85]{2009xxNOFA_Bha}). While these analogs show an equality, our generalization is at the cost of a weakening of the equality to an inequality.

\begin{figure}[tb]
 \centering
 \includegraphics[width=.2\linewidth]{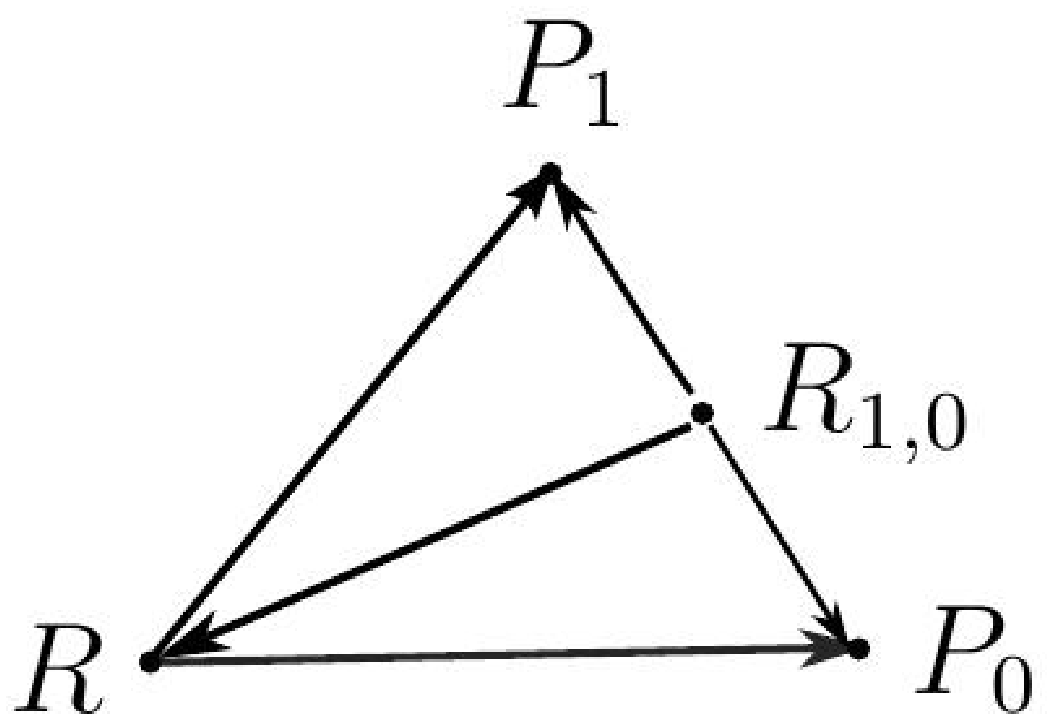}
 \caption{\label{p1:fig:parallelogram}The usual Apollonius theorem would be, with $\lambda = \textstyle\frac{1}{2}$, $\textstyle\frac{1}{2} |\protect\overrightarrow{RP_1}|^2 + \textstyle\frac{1}{2} |\protect\overrightarrow{RP_0}|^2 = \textstyle\frac{1}{2} |\protect\overrightarrow{R_{1,0}P_1}|^2 + \textstyle\frac{1}{2} |\protect\overrightarrow{R_{1,0}P_0}|^2 + |\protect\overrightarrow{R_{1,0}R}|^2$. Here, $|\protect\overrightarrow{RP_1}|^2$ is replaced by the asymmetric $I_f(P_1',R')$, etc., and the equality by an inequality whose direction depends on $\alpha <1$ or $\alpha >1$.}
\end{figure}

\begin{proposition}[Extension of Apollonius Theorem]
\label{p1:parallel}
Let $\alpha < 1$. Let $P_0, P_1, R$ be probability measures that are absolutely continuous with respect to $\mu$, and let the corresponding Radon-Nikodym derivatives $p_0, p_1,$ and $r$ be in $L^{\alpha}(\mu)$. Assume $0 \le \lambda \le 1$. We then have
\begin{eqnarray}
\label{p1:eqn:parallelogram}
 \lambda I_f(P'_1,R') + (1-\lambda ) I_f(P'_0,R') - \lambda  I_f(P'_1,R'_{1,0}) - (1-\lambda ) I_f(P'_0,R'_{1,0}) \ge I_f(R'_{1,0},R'),
\end{eqnarray}
where
\begin{equation}
  \label{p1:eqn:rstar}
  R_{1,0} = \displaystyle \frac{\frac{\lambda}{\|p_1\|}P_1+\frac{1-\lambda}{\|p_0\|}P_0}{\frac{\lambda}{\|p_1\|}+\frac{1-\lambda}{\|p_0\|}}.
\end{equation}
When $\alpha > 1$, the reversed inequality holds in (\ref{p1:eqn:parallelogram}).
\end{proposition}

\vspace*{.1in}

\begin{IEEEproof}
See Figure \ref{p1:fig:parallelogram} for an interpretation of (\ref{p1:eqn:parallelogram}) as an analog of the Apollonius Theorem. We first recognize that
\begin{eqnarray}
  \label{p1:eqn:If_p'q'}
  I_f(P', Q') & = & \text{sgn}(\rho) \left[\int \frac{p}{\|p\|} \left( \frac{q}{\|q\|} \right)^{\alpha-1} d\mu - 1\right].
\end{eqnarray}
Let $r_{1,0} = dR_{1,0}/d\mu$. Using (\ref{p1:eqn:If_p'q'}), the left-hand side of (\ref{p1:eqn:parallelogram}) can be expanded to
\begin{eqnarray*}
  \lefteqn{ \text{sgn}(\rho) \int \frac{\lambda p_1}{\|p_1\|}
                                  \left[ \left( \frac{r}{\|r\|} \right)^{\alpha-1}
                                         - \left( \frac{r_{1,0}}{\| r_{1,0} \|} \right)^{\alpha-1}
                                  \right] d\mu } \\
  & & \hspace*{-.1in}+ \text{sgn}(\rho) \int \frac{(1-\lambda) p_0}{\|p_0\|}
                                  \left[ \left( \frac{r}{\|r\|}\right)^{\alpha-1}
                                         \hspace*{-.1in} - \left( \frac{r_{1,0}}{\| r_{1,0} \|} \right)^{\alpha-1}
                                  \right] d\mu \\
  & \stackrel{(a)}{=} & \text{sgn}(\rho) \int \frac{r_{1,0}}{\| r_{1,0} \|}
                                 \left[ \left( \frac{r}{\|r\|} \right)^{\alpha-1}
                                        - \left( \frac{r_{1,0}}{\| r_{1,0} \|} \right)^{\alpha-1}
                                 \right] d \mu\\
  & & \times \left[ \frac{\lambda}{\|p_1\|}+\frac{1-\lambda}{\|p_0\|} \right] \| r_{1,0}\| \\
  & \stackrel{(b)}{=} & \left[ \frac{\lambda}{\|p_1\|}+\frac{1-\lambda}{\|p_0\|} \right] \| r_{1,0}\| \cdot I_f(R'_{1,0}, R'),
\end{eqnarray*}
where (a) follows from (\ref{p1:eqn:rstar}) and after a multiplication and a division by the scalar $\| r_{1,0} \|$; (b) follows from (\ref{p1:eqn:If_p'q'}). The lemma would follow if we can show
\[
  \left(\frac{\lambda}{\|p_1\|}+\frac{1-\lambda}{\|p_0\|}\right) \|r_{1,0}\| \ge 1
\]
for $\alpha < 1$, and the reversed inequality for $\alpha > 1$. But these are direct consequences of Minkowski's inequalities for $\alpha < 1$ and $\alpha > 1$ applied to (\ref{p1:eqn:rstar}).
\end{IEEEproof}

\vspace*{.1in}

Let us now formally define what we mean by a forward $\mathscr{I}_{\alpha}$-projection.

\vspace*{.1in}

\begin{definition}
  If $\mathbb{E}$ is a set of probability measures on $(\mathbb{X},\mathcal{X})$ such that $\mathscr{I}_{\alpha}(P,R)<\infty$ for some $P\in \mathbb{E}$, a measure $Q \in \mathbb{E}$ satisfying
  \begin{eqnarray}
    \label{p1:eqn:projection}\mathscr{I}_{\alpha}(Q,R)=\displaystyle \inf_{P\in \mathbb{E}}\mathscr{I}_{\alpha}(P,R)=: \mathscr{I}_{\alpha}(\mathbb{E},R)
  \end{eqnarray}
  is called a forward \emph{$\mathscr{I}_{\alpha}$-projection} of $R$ on $\mathbb{E}$.
\end{definition}

\vspace*{.1in}

For a set $\mathbb{E}$ of probability measures on $(\mathbb{X}, \mathcal{X})$, let
\[
  \mathcal{E} := \left\{ p = \frac{dP}{d\mu} \colon P \in \mathbb{E} \right\}
\]
be the corresponding set of $\mu$-densities. We shall assume that $\mathcal{E} \subset L^{\alpha}(\mu)$.

We are now ready to state our first main result on the existence and uniqueness of the forward $\mathscr{I}_{\alpha}$-projection.

\vspace*{.1in}

\begin{thm}[Existence and uniqueness of the forward $\mathscr{I}_{\alpha}$-projection]
\label{p1:thm:projection}
Fix $\alpha > 0$, $\alpha \neq 1$. Let $\mathbb{E}$ be a set of probability measures whose corresponding set of density functions $\mathcal{E}$ is convex and closed in $L^{\alpha}(\mu)$. Let $R$ be a probability measure (with density $r$) and suppose  that $\mathscr{I}_{\alpha}(P,R) < \infty$ for some $P \in \mathbb{E}$. Then $R$ has a unique forward $\mathscr{I}_{\alpha}$-projection on $\mathbb{E}$.
\end{thm}

\vspace*{.1in}

\begin{remark}
This is a generalization of Csisz\'{a}r's projection result \cite[Th.~2.1]{1975xxAP_Csi} for relative entropy ($\alpha = 1$). The analog of ``$\mathcal{E}$ is closed in $L^{\alpha}(\mu)$'' for relative entropy is closure in the topology arising from the total variation metric, one of the hypotheses in \cite[Th.~2.1]{1975xxAP_Csi}. The proof ideas are different for the two cases $\alpha < 1$ and $\alpha > 1$. The proof for $\alpha < 1$ is a modification of Csisz\'{a}r's approach in \cite{1975xxAP_Csi}, and is similar to the classical proof of existence and uniqueness of the {\em best approximant} of a point (in a Hilbert space) from a given closed and convex set of the Hilbert space. (See, for e.g.,  \cite[Ch.~11,~Th.~14]{2009xxNOFA_Bha}). The proof for $\alpha > 1$ exploits the reflexive property of the Banach space $L^{\alpha}(\mu)$. This alternative approach is required because the inequality in the extension of Apollonius Theorem (Proposition \ref{p1:parallel}) is in a direction that renders the classical approach inapplicable. We are indebted to Pietro Majer for suggesting some key steps for the $\alpha > 1$ case on the {\tt mathoverflow.net} forum.
\end{remark}

\vspace*{.1in}

\begin{remark}
In general, when $\alpha \neq 1$, the forward $\mathscr{I}_{\alpha}$-projection depends on the reference measure $\mu$. The case $\alpha = 1$ of relative entropy is however special in that the forward $\mathscr{I}_1$-projection does not depend on the reference measure $\mu$.
\end{remark}

\vspace*{.1in}

\begin{remark}
The above result was established by Sundaresan \cite[Prop.~23]{200701TIT_Sun} for finite $\mathbb{X}$. That proof relied on the compactness of $\mathbb{E}$. The current proof works for general measure spaces.
\end{remark}

\vspace*{.1in}

\begin{IEEEproof}
(a) We first consider the case $\alpha < 1$.

{\em Existence of forward projection}: Pick a sequence $(P_n)$ in $\mathbb{E}$ such that $I_f(P'_n,R') < \infty$ and
\begin{eqnarray}
  \label{p1:eqn:inf}
  I_f(P_n',R')\to\displaystyle\inf_{P\in \mathbb{E}}I_f(P',R').
\end{eqnarray}
Apply Proposition \ref{p1:parallel} with $\lambda = \frac{1}{2}$ to get
\begin{eqnarray}
\label{p1:eqn:projectionProofStep1}
  \textstyle\frac{1}{2} I_f(P_m',R') + \textstyle\frac{1}{2} I_f(P_n',R') - \textstyle\frac{1}{2} I_f(P_m',R'_{m,n}) - \textstyle\frac{1}{2} I_f(P_n',R'_{m,n}) \ge ~ I_f(R'_{m,n},R'),
\end{eqnarray}
where
\[
  R_{m,n} =\displaystyle\frac{\frac{1}{\|p_m\|}P_m + \frac{1}{\|p_n\|}P_n}
                             {\frac{1}{\|p_m\|}+\frac{1}{\|p_n\|}}.
\]
$R_{m,n}  \in \mathbb{E}$ on account of the convexity of $\mathbb{E}$. Using $I_f(\cdot,\cdot) \geq 0$ and then rearranging (\ref{p1:eqn:projectionProofStep1}), we get
\begin{eqnarray}
  \label{p1:eqn:projectionProofStep2}
  0  & \le & \textstyle\frac{1}{2} I_f(P'_m, R'_{m,n}) + \textstyle\frac{1}{2} I_f(P'_n,R'_{m,n}) \\ \label{p1:eqn:projectionProofStep3}
        & \le & \textstyle\frac{1}{2} I_f(P'_m, R') + \textstyle\frac{1}{2} I_f(P'_n, R') - I_f(R'_{m,n}, R').
\end{eqnarray}
Now let $m,n \to \infty$. We claim the expression on the right-hand side of (\ref{p1:eqn:projectionProofStep3}) must approach 0. Indeed, that the liminf of the right-hand side of (\ref{p1:eqn:projectionProofStep3}) is at least 0 is clear from the inequalities (\ref{p1:eqn:projectionProofStep2}) and (\ref{p1:eqn:projectionProofStep3}). But the limsup is at most 0 because both $I_f(P'_m, R')$ and $I_f(P'_n, R')$ approach the infimum value, and $I_f(R'_{m,n}, R')$ is at least this infimum value for each $m$ and $n$. This establishes the claim.

Consequently, the right-hand side of (\ref{p1:eqn:projectionProofStep2}) converges to 0. Using this and the nonnegativity of $I_f(\cdot,\cdot)$, we get
\begin{equation}
  \label{p1:eqn:Iflim}
  \lim_{m,n \to \infty} I_f(P_m',R'_{m,n}) = 0.
\end{equation}
From \cite[Th.~1]{1967xxSSMH2_Csi}, a generalization of Pinsker's inequality for $f$-divergence under $\alpha < 1$, and with $|P - Q|_{TV}$ denoting the total variation distance between probability measures $P$ and $Q$, we have
\[
  \lim_{m,n \to \infty} | P'_m - R'_{m,n} |_{TV} = 0.
\]
The triangle inequality for the total variation metric then yields
\[
  |P_m'-P_n'|_{TV} \le |P_n'-R'_{m,n}|_{TV} + |P_m'-R'_{m,n}|_{TV} \to 0
\]
as $m,n \to \infty$, i.e., the sequence $(p_n')$ is a Cauchy sequence in $L^1(\mu)$. It must therefore converge to some $g$ in $L^1(\mu)$, i.e.,
\begin{equation}
  \label{p1:eqn:pn/normpn}
    \lim_{n \to \infty} \int \left| p_n' - g \right| d\mu = 0.
\end{equation}
It follows that $\int p_n' d\mu \rightarrow \int g d\mu$, and since $\int p_n' d\mu = 1$ for all $n$, we must have $\int g d\mu = 1$.

From the $L^1(\mu)$ convergence in (\ref{p1:eqn:pn/normpn}), we also have $p_n' \to g$ in $[\mu]$-measure.

We will now demonstrate that the probability measure with $\mu$-density proportional to $g^{1/\alpha}$ is in $\mathbb{E}$ and is a forward $\mathscr{I}_{\alpha}$-projection, thereby establishing existence.

In view of the convergence in $[\mu]$-measure and the upper bound
\[
  \left| (p_n')^{1/\alpha} - g^{1/\alpha} \right|^{\alpha} \le 2^{\alpha} \left[ p_n' + g \right],
\]
we can apply the generalized version of the dominated convergence theorem (\cite[Ch.~2,~Ex.~20]{1999xxRA_Fol} or \cite[p.139, Problem 19]{2001xxLIES_Jon}) to get
\[
  \frac{p_n}{\|p_n\|} = (p_n')^{1/\alpha} \to g^{1/\alpha} \mbox{ in } L^{\alpha}(\mu).
\]

We next claim that
\begin{equation}
  \label{p1:eqn:norm-pn-bounded}
  \|p_n\| \mbox{ is bounded.}
\end{equation}
Suppose not; then working on a subsequence if needed, we have $\|p_n\| := M_n \to \infty$. As $\int p_n d\mu=1$, given any $\epsilon>0$,
\[
  \mu \left( \left\{ p_n'>\epsilon \right\} \right)
    = \mu \left( \left\{ p_n>\epsilon^{1/\alpha} M_n \right\} \right)
    \le \frac{1}{\epsilon^{1/\alpha} M_n}\to 0 \mbox{ as } n\to \infty,
\]
and hence $p_n'\to 0$ in $[\mu]$-measure, or $g = 0$ except on a set of $[\mu]$-measure 0 (i.e., $g=0$ a.e.$[\mu]$) . But this is a contradiction since $\int g \, d\mu=1$. Thus (\ref{p1:eqn:norm-pn-bounded}) holds, and we can pick a subsequence of the sequence $(\|p_n\|)$ that converges to some $c$. Reindex and work on this subsequence to get $p_n \to cg^{1/\alpha}$ in $L^{\alpha}(\mu)$.

It is now that we use the hypothesis that $\mathcal{E}$ is closed in $L^{\alpha}(\mu)$. We remind the reader that $\mathcal{E}$ is the set of $\mu$-densities of members of $\mathbb{E}$. The closedness implies that the limiting function $cg^{1/\alpha}=q$ for some $q \in \mathcal{E}$, and so $q$ must be the density of a probability measure, say $Q$. Since we also have $\int g d\mu = 1$, it follows that $c=\|q\|$ and $g = q^{\alpha}/\|q\|^{\alpha}$. As $p_n \to q$ in $L^{\alpha}(\mu)$, lower semicontinuity of $\mathscr{I}_{\alpha}(\cdot,r)$ (Proposition \ref{p1:prop:lsc1}) implies
\begin{eqnarray}
  \label{p1:eqn:liminf}
  \mathscr{I}_{\alpha}(Q,R) \le \liminf_{n \to \infty} \mathscr{I}_{\alpha}(P_n, R) = \mathscr{I}_{\alpha}(\mathbb{E}, R).
\end{eqnarray}
Since $Q\in \mathbb{E}$, $\mathscr{I}_{\alpha}(Q,R) \ge \mathscr{I}_{\alpha}(\mathbb{E}, R)$, and therefore equality must hold in (\ref{p1:eqn:liminf}), and $Q$ is a forward $\mathscr{I}_{\alpha}$-projection of $R$ on $\mathbb{E}$.

{\em Uniqueness}: Our proof of uniqueness is analogous to the usual proof of uniqueness of projection in Hilbert spaces \cite[p.~86]{2009xxNOFA_Bha}. A simpler proof, after the `Pythagorean property' is established, can be found at the end of Section \ref{p1:sec:pythagorenproperty}.

Write $d$ for the infimum value in the right-hand side of (\ref{p1:eqn:inf}) and let $Q_1$ and $Q_0$ attain the infimum. Apply Proposition \ref{p1:parallel} with $\lambda = 1/2$ and with $Q_1$ and $Q_0$ in place of $P_1$ and $P_0$ to get
\begin{equation}
 \label{p1:eqn:projectionunique}
\textstyle\frac{1}{2} I_f(Q'_1,R') + \textstyle\frac{1}{2} I_f(Q'_0,R') - \textstyle\frac{1}{2} I_f(Q'_1,R'_{1,0}) - \textstyle\frac{1}{2} I_f(Q'_0,R'_{1,0})  \ge I_f(R'_{1,0},R'),
\end{equation}
where
\[
  R_{1,0} = \displaystyle \frac{\frac{1}{\|q_1\|}Q_1+\frac{1}{\|q_0\|}Q_0}{\frac{1}{\|q_1\|}+\frac{1}{\|q_0\|}}.
\]
Since $R_{1,0} \in \mathbb{E}$ we have $I_f(R'_{1,0},R') \ge d$. Use this in (\ref{p1:eqn:projectionunique}), substitute $I_f(Q_i',R') = d,~i=0,1$, and we get
\begin{eqnarray*}
\textstyle\frac{d}{2} + \textstyle\frac{d}{2} - \textstyle\frac{1}{2} I_f(Q_1',R_{1,0}') - \textstyle\frac{1}{2} I_f(Q_0',R_{1,0}') \geq d,
\end{eqnarray*}
and this implies
\[
  I_f(Q_1',R_{1,0}') + I_f(Q_0',R_{1,0}') \leq 0.
\]
The nonnegativity of each of the terms then implies that each must be zero, and so $Q_1 = R_{1,0} = Q_0$. The forward $\mathscr{I}_{\alpha}$-projection is unique.

This completes the proof for the case when $\alpha < 1$.

\vspace*{.1in}

(b) We now consider the case when $\alpha > 1$.

{\em Existence of forward projection}: Equation (\ref{p1:eqn:projection}) can be rewritten (using (\ref{p1:alphadiv_linear_form_general})) as
\begin{eqnarray}
  \label{p1:eqn:inftosup}
  \inf_{P \in \mathbb{E}} \mathscr{I}_{\alpha}(P,R)
    & = & \frac{1}{\rho} \log \left[ \sup_{p \in \mathcal{E}} \int \frac{p}{\|p\|} \left( \frac{r}{\|r\|} \right)^{\alpha-1} d\mu \right] \\
    \label{p1:eqn:blf}
    & = & \frac{1}{\rho} \log \left[ \sup_{h \in \hat{\mathcal{E}}} \int h g \, d\mu \right],
\end{eqnarray}                                                                                                                                                                                                                                                                                                                                                                                                       where
$$\hat{\mathcal{E}} := \left\{ s \frac{p}{\|p\|} \colon p \in \mathcal{E}, 0 \le s\le 1 \right\},$$
and $g=\left(r / \|r\| \right)^{\alpha-1}$, an element of the dual space $\left(L^{\alpha}(\mu)\right)^*$. Allowing $s \in [0,1]$ makes $\hat{\mathcal{E}}$ convex (as we shall soon show), but does not change the supremum.

We now claim that
\begin{equation}
  \label{p1:eqn:claim-closed-convex}
  \hat{\mathcal{E}} \mbox{ is a closed and convex subset of } L^{\alpha}(\mu).
\end{equation}

Assume the claim. Since $L^{\alpha}(\mu)$ is a reflexive Banach space for $\alpha > 1$, the convex and closed set $\hat{\mathcal{E}}$ is also closed in the {\em weak topology} \cite[Ch.~10,~Cor.~23]{1988xxRA_Roy}. Using the Banach-Alaoglu theorem and the fact that $L^{\alpha}(\mu)$ is a reflexive Banach space, we have that the unit ball is compact in the weak topology. Since $\hat{\mathcal{E}}$ is a (weakly) closed subset of a (weakly) compact set, $\hat{\mathcal{E}}$ is (weakly) compact. The linear functional $h \mapsto \int h g \, d\mu$ is continuous in the weak topology, and hence the supremum over the (weakly) compact set $\hat{\mathcal{E}}$ is attained. Since the linear functional increases with $s$, the supremum is attained when $s=1$, i.e., there exists a $p \in \mathcal{E}$ for which the supremum in (\ref{p1:eqn:inftosup}) is attained.

We now proceed to show the claim (\ref{p1:eqn:claim-closed-convex}). To see convexity, let $p_1, p_0 \in \mathcal{E}$, let $0 \le s_1, s_0 \le 1$, and let $0 \le \lambda \le 1$. The convex combination of $s_1 p_1 / \|p_1\|$ and $s_0 p_0 / \| p_0 \|$ is
\[
  \lambda s_1 \frac{p_1}{\|p_1\|} + (1-\lambda) s_0 \frac{p_0}{\|p_0\|}.
\]
If both $\lambda s_1$ and $(1-\lambda)s_0$ are zero, then this convex combination is 0 which is trivially in $\hat{\mathcal{E}}$. Otherwise, we can write the convex combination as
\begin{equation}
  \label{p1:eqn:h-mix}
  \lambda s_1 \frac{p_1}{\|p_1\|} + (1-\lambda) s_0 \frac{p_0}{\|p_0\|}
    = s_{\lambda} \frac{p_{\lambda}}{\|p_{\lambda}\|},
\end{equation}
where
\begin{eqnarray}
  \label{p1:eqn:plambda}
  p_{\lambda} & := & \frac{\frac{\lambda s_1}{\|p_1\|}p_1 + \frac{(1-\lambda)s_0}{\|p_0\|}p_0}
            {\frac{\lambda s_1}{\|p_1\|} + \frac{(1-\lambda)s_0}{\|p_0\|}}, \\
  \label{p1:eqn:slambda}
  s_{\lambda} & := & \left( \frac{\lambda s_1}{\|p_1\|} + \frac{(1-\lambda)s_0}{\|p_0\|} \right) \cdot \|p_{\lambda}\|.
\end{eqnarray}
To show that the convex combination is in $\hat{\mathcal{E}}$, it suffices to show that $p_{\lambda} \in \mathcal{E}$ and $s_{\lambda} \in [0,1]$.

The convexity of $\mathcal{E}$ immediately implies that $p_{\lambda} \in \mathcal{E}$. It is also clear that $s_{\lambda} \geq 0$. From Minkowski's inequality (for $\alpha > 1$), we have
\begin{eqnarray}
  s_{\lambda} & = & \left( \frac{\lambda s_1}{\|p_1\|} + \frac{(1-\lambda)s_0}{\|p_0\|} \right) \cdot \|p_{\lambda}\| \nonumber \\
    & = & \left\| \frac{\lambda s_1}{\|p_1\|}p_1 + \frac{(1-\lambda)s_0}{\|p_0\|}p_0 \right\| \nonumber \\
    & \le & \frac{\lambda s_1}{\|p_1\|} \cdot \| p_1 \| + \frac{(1-\lambda)s_0}{\|p_0\|} \cdot \| p_0 \| \nonumber \\
    & = & \lambda s_1 + (1- \lambda) s_0 \nonumber \\
    & \leq & 1. \label{p1:eqn:Minkowski}
\end{eqnarray}
This establishes that $\hat{\mathcal{E}}$ is convex.

To see that $\hat{\mathcal{E}}$ is closed in $L^{\alpha}(\mu)$, let $(g_n)$ be a sequence in $\hat{\mathcal{E}}$ such that $g_n \to g$ for some $g \in L^{\alpha}(\mu)$. We need to show $g \in \hat{\mathcal{E}}$.

Write $g_n = s_n p_n / \|p_n\|$, where $p_n \in \mathcal{E}$ and $0 \leq s_n \leq 1$. Since $g_n \to g$ in $L^{\alpha}(\mu)$, take norms to get $s_n = \| g_n \| \to \| g \|$, and so $\|g\| \leq 1$.

If $\| g \| = 0$, then $g=0$ a.e.$[\mu]$, and so $g$ trivially belongs to $\hat{\mathcal{E}}$. We may therefore assume $\| g \| > 0$. It follows that $p_n / \|p_n\| = g_n / \|g_n\| \to g / \|g\|$ in $L^{\alpha}(\mu)$.

Again, as in (\ref{p1:eqn:norm-pn-bounded}), we claim that $\| p_n \|$ is bounded. Suppose not. As in the proof of (\ref{p1:eqn:norm-pn-bounded}), move to a subsequence if needed and assume $\| p_n \| := M_n \to \infty$. As $\int p_n d\mu = 1$, we have
\[
  \mu \left( \left\{ \frac{p_n}{\| p_n \|} > \epsilon \right\} \right) = \mu \left( \left\{ p_n > \epsilon M_n \right\} \right) \leq \frac{1}{\epsilon M_n} \to 0
\]
as $n \to \infty$, and $p_n / \| p_n \| \to 0$ in $\mu$-measure, or its limit $g / \|g\| = 0$ a.e.$[\mu]$. But this contradicts the fact that $\int \left( g / \| g \| \right)^{\alpha} d\mu = 1$. Thus $\| p_n \|$ is bounded.

Focusing on a subsequence, if needed, we may assume $\| p_n \| \to c$ for some $c \geq 0$. Hence $p_n \to cg/\|g\|$ in $L^{\alpha}(\mu)$. Since $\mathcal{E}$ is closed, we must have $cg/\|g\|=p$ for some $p\in \mathcal{E}$, whence $c=\|p\|$ and $g=\|g\| \cdot p / \|p\|$. Since we already established that $\|g\| \leq 1$, it follows that $g \in \hat{\mathcal{E}}$.

{\em Uniqueness}: We now proceed to show uniqueness.

Let $p_0, p_1$ attain the supremum in (\ref{p1:eqn:inftosup}). Set $h_0 = s_0 p_0/\| p_0 \|$ and $h_1 = s_1 p_1 / \| p_1 \|$ with $s_0 = s_1 = 1$. Clearly $h_0$ and $h_1$ attain the supremum in (\ref{p1:eqn:blf}). By convexity of $\hat{\mathcal{E}}$, $\frac{1}{2}h_1 + \frac{1}{2}h_0$ belongs to $\hat{\mathcal{E}}$. This and the linearity of the integral in (\ref{p1:eqn:blf}) in the $h$ variable imply that $\frac{1}{2}h_1 + \frac{1}{2}h_0$ attains the supremum in (\ref{p1:eqn:blf}). Noticing that $\frac{1}{2}h_1 + \frac{1}{2}h_0 = s_{\frac{1}{2}} p_{\frac{1}{2}}/ \| p_{\frac{1}{2}} \|$ as in (\ref{p1:eqn:h-mix}), with $p_{\frac{1}{2}}$ and $s_{\frac{1}{2}}$ as in (\ref{p1:eqn:plambda}) and (\ref{p1:eqn:slambda}), respectively, we gather that $s_{\frac{1}{2}} = 1$. Consequently, all the inequalities in the chain (\ref{p1:eqn:Minkowski}) must be equalities. But then $p_1$ and $p_0$ are scalings of each other (which is the condition for equality in Minkowski's inequality). Since $p_0$ and $p_1$ are densities of probability measures with respect to $\mu$, we deduce that the scaling factor must be 1, i.e., $p_0 = p_1$. This completes the proof.
\end{IEEEproof}

%
%
%
%
%
%

\section{Pythagorean property}
\label{p1:sec:pythagorenproperty}

\begin{figure}[tb]
\centering
\begin{minipage}{.4\textwidth}
  \centering
  \includegraphics[width=.42\linewidth]{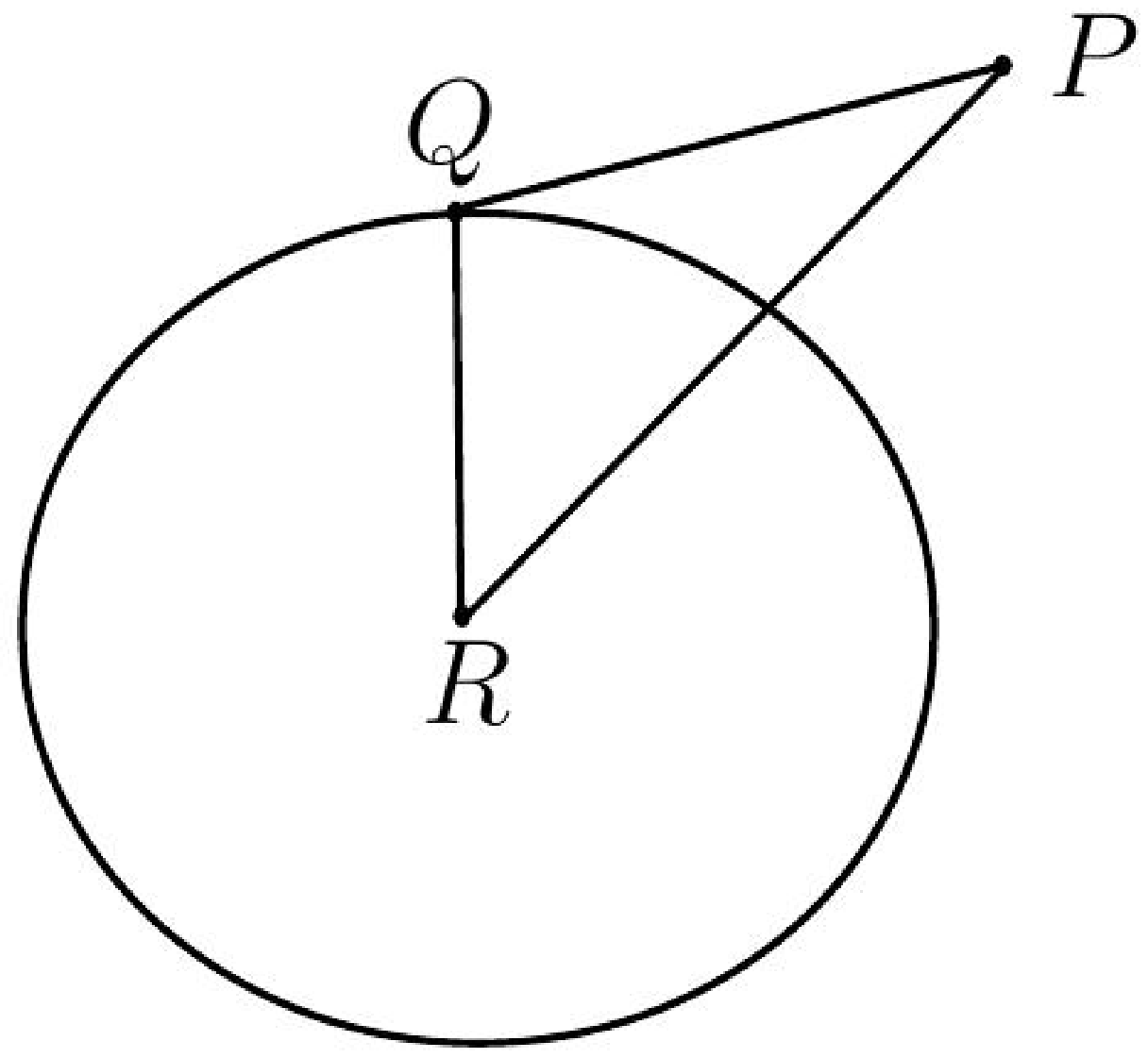}
  \captionof{figure}{Pythgorean property with inequality}
  \label{fig:pyth(a)}
\end{minipage}
\begin{minipage}{.475\textwidth}
  \centering
  \includegraphics[width=.45\linewidth]{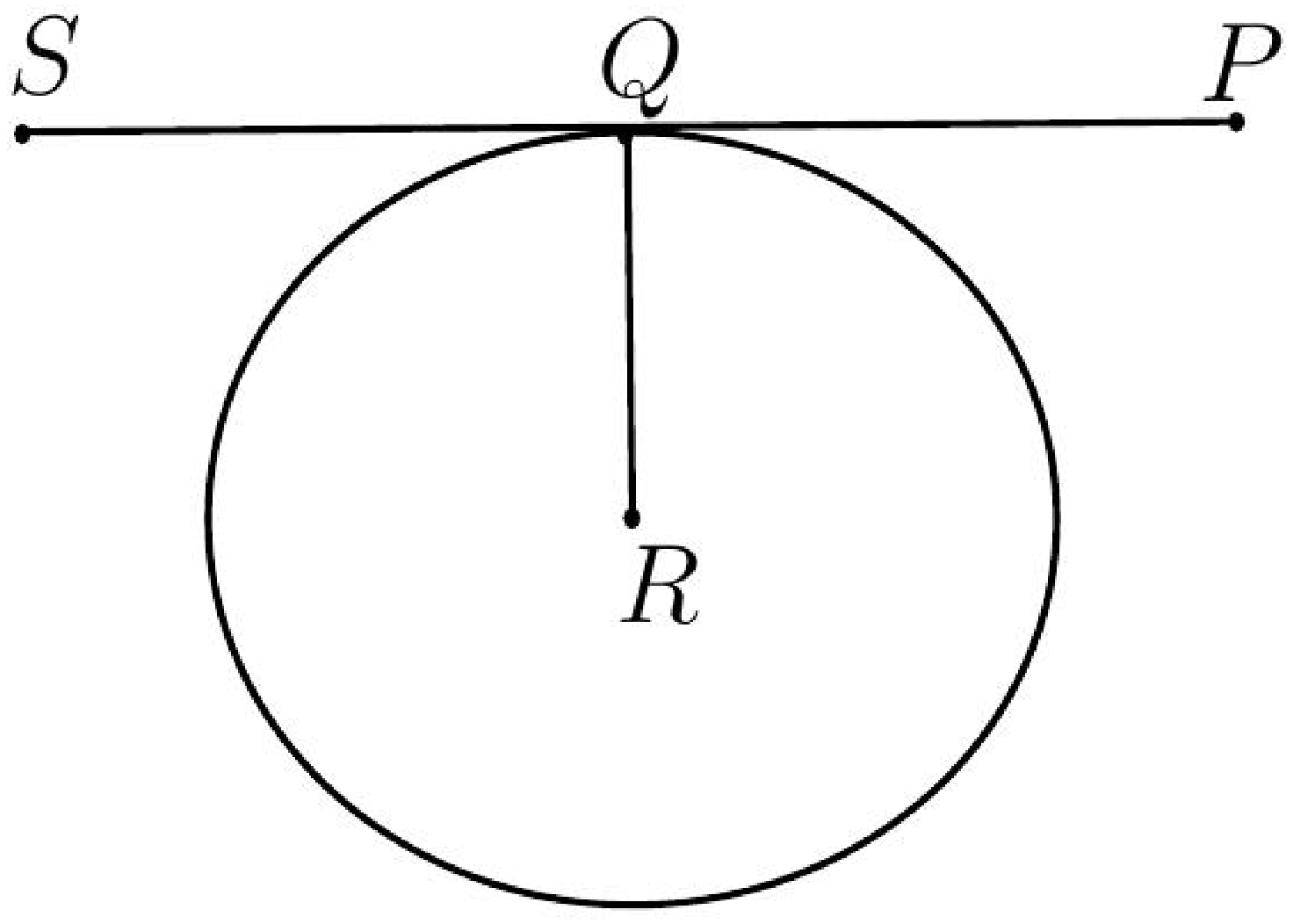}
  \captionof{figure}{Pythgorean property with equality}
  \label{fig:pyth(b)}
\end{minipage}
\end{figure}

In this section, we state and prove the Pythagorean property for relative $\alpha$-entropy. We define the $\mathscr{I}_{\alpha}$-ball with center $R$ and radius $\tau$ to be $B(R,\tau) := \{P\colon\mathscr{I}_{\alpha}(P,R)<\tau\},~ 0<\tau\le \infty$. By virtue of quasi-convexity, $B(R,\tau)$ is a convex set.

\vspace*{.1in}

\begin{thm}[The Pythagorean property]
\label{p1:thm:pythagorean}
Let $\alpha>0$ and $\alpha\neq 1$.
\begin{itemize}
\item[(a)] Let $\mathscr{I}_{\alpha}(P,R)$ and $\mathscr{I}_{\alpha}(Q,R)$ be finite. The segment joining $P$ and $Q$ does not intersect the $\mathscr{I}_{\alpha}$-ball $B(R,\tau)$ with radius $\tau=\mathscr{I}_{\alpha}(Q,R)$, i.e., $\mathscr{I}_{\alpha}(P_{\lambda},R)\ge \mathscr{I}_{\alpha}(Q,R)$ for
    \[
      P_{\lambda}=\lambda P+(1-\lambda)Q,~\lambda\in [0,1],
    \]
    (see figure \ref{fig:pyth(a)}) if and only if
    \begin{eqnarray}
      \label{p1:eqn:pythagorean_inequality}
      \mathscr{I}_{\alpha}(P,R)\ge \mathscr{I}_{\alpha}(P,Q)+\mathscr{I}_{\alpha}(Q,R).
    \end{eqnarray}
\item[(b)] Let
    \begin{eqnarray}
      \label{p1:eqn:convex combination} Q = \lambda P+(1-\lambda)S, \, \text{for some fixed} \, \lambda\in (0,1),
    \end{eqnarray}
    and let $\mathscr{I}_{\alpha}(Q,R)$ be finite. The segment joining $P$ and $S$ does not intersect $B(R,\tau)$ with $\tau=\mathscr{I}_{\alpha}(Q,R)$ (see figure \ref{fig:pyth(b)}) if and only if the following two equalities hold:
  \begin{equation}
     \label{p1:eqn:pythagorean_equality}
     \left.
     \begin{array}{ccc}
     \mathscr{I}_{\alpha}(P,R) & = & \mathscr{I}_{\alpha}(P,Q)+\mathscr{I}_{\alpha}(Q,R) \\
     \mathscr{I}_{\alpha}(S,R) & = & \mathscr{I}_{\alpha}(S,Q)+\mathscr{I}_{\alpha}(Q,R).
     \end{array}
     \right\}
  \end{equation}
 \end{itemize}
\end{thm}


\vspace*{.1in}

\begin{IEEEproof}
Our proof proceeds as in \cite{200701TIT_Sun}, where the above result is proved for the finite alphabet case, with appropriate functional analytic justifications to account for the generality of the alphabet.

(a) We begin with the ``only if'' part. Assume $\mathscr{I}_{\alpha}(P,R)$ and $\mathscr{I}_{\alpha}(Q,R)$ are finite, and that the segment joining $P$ and $Q$ does not intersect the $\mathscr{I}_{\alpha}$-ball $B(R,\tau)$ with radius $\tau=\mathscr{I}_{\alpha}(Q,R)$. To show (\ref{p1:eqn:pythagorean_inequality}), since
\begin{eqnarray*}
  I_{f}(P',R')
    & = & \int r'f\left(\frac{p'}{r'}\right)d\mu\\
    & = & \text{sgn}(\rho) \left[\int (p')^{1+\rho}\cdot (r')^{-\rho}d\mu - 1\right]\\&=& \text{sgn}(\rho) \left[\int \frac{p}{\|p\|} \cdot(r')^{-\rho} d\mu - 1\right],
\end{eqnarray*}
which follow from (\ref{p1:eqn:escort}), (\ref{p1:eqn:If-defn}), and $\alpha (1+\rho)$, it suffices to show that
\begin{eqnarray}
\label{p1:eqn:equivalentpythagoreanidentity}
  \text{sgn}(\rho)\int p \cdot (r')^{-\rho} d\mu \ge \frac{\text{sgn}(\rho)}{\|q\|}\int p \cdot (q')^{-\rho} d\mu \cdot \int q \cdot (r')^{-\rho} d\mu.
\end{eqnarray}
We have
\begin{eqnarray*}
  I_{f}(P_{\lambda}',R')
    & = & \text{sgn}(\rho) \left[\int \frac{p_{\lambda}}{\|p_{\lambda}\|}\cdot (r')^{-\rho} d\mu - 1\right]
\end{eqnarray*}
Let
\begin{eqnarray*}
  s(\lambda) & := & \int p_{\lambda}\cdot (r')^{-\rho}d\mu, \\
  t(\lambda) & := & \|p_{\lambda}\|.
\end{eqnarray*}
Clearly, $\mathscr{I}_{\alpha}(P_{\lambda},R)\ge \mathscr{I}_{\alpha}(Q,R)$ for $\lambda \in (0,1)$ implies that
\begin{eqnarray} \label{p1:eqn:derivativeoff-divergence}\frac{I_{f}(P_{\lambda}',R')-I_{f}(P_{0}',R')}{\lambda}\ge 0 ~~\text{for}~~ \lambda \in (0,1).\end{eqnarray}
Therefore, by taking the limit as $\lambda \rightarrow 0$, the derivative of $I_f(P_{\lambda}',R')$ with respect to $\lambda$ evaluated at $\lambda=0$, should be $\ge 0$. Observe that
\begin{eqnarray*}
  \frac{s(\lambda)-s(0)}{\lambda}
   & = & \frac{1}{\lambda}\left[\int p_{\lambda}\cdot (r')^{-\rho}d\mu-\int q \cdot (r')^{-\rho}d\mu\right]\\
   & = & \int \left(\frac{p_{\lambda}-q}{\lambda}\right) \cdot (r')^{-\rho}d\mu\\
   & = & \int (p-q) \cdot (r')^{-\rho}d\mu\\ & = & \left[\int p \cdot (r')^{-\rho}d\mu-\int q \cdot (r')^{-\rho}d\mu\right].
\end{eqnarray*}
So $\dot{s}(0) := \lim_{\lambda \downarrow 0} (s(\lambda)-s(0))/\lambda$ exists and equals the above expression.

Let us now identify $\dot{t}(0)$. For $\alpha>1$ (i.e., $\rho > 0$), we have
\[
  \left| \frac{\partial{}}{\partial \lambda}(p_{\lambda})^{\alpha} \right| = \alpha(p_\lambda)^{\alpha-1} |p-q|\le \alpha (p+q)^{\alpha},
\]
while for $\alpha<1$, notice that for any $0 < l < \frac{1}{2}$, we have
\begin{eqnarray*}
  \left| \frac{\partial{}}{\partial \lambda}(p_{\lambda})^{\alpha} \right|
    & = & \alpha (p_\lambda)^{\alpha-1} |p-q|
    = \frac{\alpha |p-q|}{[\lambda p +(1-\lambda)q]^{1-\alpha}}
    \le \frac{\alpha (p+q)}{\min\{\lambda , (1-\lambda)\}^{1-\alpha}(p+q)^{1-\alpha}}\\
    & \le & \frac{\alpha (p+q)^{\alpha}}{l^{1-\alpha}} \hspace*{.1in} \forall \lambda\in (l,1-l),
  \end{eqnarray*}
and both upper bounds are in $L^1(\mu)$. Therefore by chain rule and \cite[Th.~2.27]{1999xxRA_Fol}, we get
\[
  \dot{t}(\lambda) = \left[ \int (p_{\lambda})^{\alpha} d\mu \right]^{\frac{1}{\alpha}-1} \cdot \int (p_{\lambda})^{\alpha-1}(p-q) d\mu
\]
for $\lambda \in (l,1-l)$. As $\lambda \downarrow 0$ (by moving $l$ closer to $0$), we get
\begin{eqnarray*}
  \dot{t}(0)
    & = & \left(\int q^{\alpha}d\mu\right)^{\frac{1}{\alpha}-1}\cdot\int q^{\alpha-1}(p-q)d\mu \\
    & = & \left(\int q^{\alpha}d\mu\right)^{\frac{1-\alpha}{\alpha}}\cdot\left(\int pq^{\alpha-1}d\mu-\int q^{\alpha}d\mu\right) \\
    & = & \int p\left(\frac{q^{\alpha}}{\int q^{\alpha}d\mu}\right)^{\frac{\alpha-1}{\alpha}}d\mu-\left(\int q^{\alpha}d\mu\right)^{\frac{1}{\alpha}} \\
    & = & \int p\cdot (q')^{-\rho}d\mu - \|q\|.
\end{eqnarray*}
Since
\begin{equation*}
  \displaystyle\frac{1}{\lambda}\left[\frac{s(\lambda)}{t(\lambda)}-\frac{s(0)}{t(0)}\right] = \frac{1}{t(\lambda)t(0)} \left[t(0)\frac{s(\lambda)-s(0)}{\lambda}-s(0)\frac{t(\lambda)-t(0)}{\lambda}\right],
\end{equation*}
it follows that the derivative of $s(\lambda)/t(\lambda)$ exists at $\lambda=0$ and is given by $(t(0)\dot{s}(0)-s(0)\dot{t}(0))/t^2(0)$. Equation (\ref{p1:eqn:derivativeoff-divergence}) and $t(0)>0$ imply that
\begin{eqnarray}
  \label{p1:eqn:nonnegativity}
  \dot{s}(0)-s(0)\cdot\frac{\dot{t}(0)}{t(0)}\ge0.
\end{eqnarray}
Consequently, $\dot{t}(0)$ is necessarily finite. Substituting the values of $s(0), \dot{s}(0), t(0)$ and $\dot{t}(0)$ in (\ref{p1:eqn:nonnegativity}) we get the required inequality (\ref{p1:eqn:equivalentpythagoreanidentity}).

To prove the converse ``if'' part, let us assume that
\[
  \mathscr{I}_{\alpha}(P,R)\ge \mathscr{I}_{\alpha}(P,Q)+\mathscr{I}_{\alpha}(Q,R),
\]
which is the same as (\ref{p1:eqn:equivalentpythagoreanidentity}). Since $\mathscr{I}_{\alpha}(P,R)$ and $\mathscr{I}_{\alpha}(Q,R)$ are finite, it follows that $\mathscr{I}_{\alpha}(P,Q)$ is also finite. From the trivial statement
\begin{equation}
 \label{p1:eqn:trivial-statement}
\mathscr{I}_{\alpha}(Q,R) = \mathscr{I}_{\alpha}(Q,Q) + \mathscr{I}_{\alpha}(Q,R),
\end{equation}
we get the following analog of (\ref{p1:eqn:equivalentpythagoreanidentity}) but with equality (replace $p$ in (\ref{p1:eqn:equivalentpythagoreanidentity}) with $q$):
\begin{eqnarray}
\label{p1:eqn:equality}
  \text{sgn}(\rho) \int q \cdot (r')^{-\rho} d\mu = \frac{\text{sgn}(\rho)}{\|q\|} \int q \cdot (q')^{-\rho} d\mu \cdot \int q \cdot (r')^{-\rho} d\mu.
\end{eqnarray}
The $\lambda$ and $(1-\lambda)$ weighted linear combination of (\ref{p1:eqn:equivalentpythagoreanidentity}) and (\ref{p1:eqn:equality}), respectively, yields,
\begin{eqnarray*}
  \text{sgn}(\rho) \int p_{\lambda} \cdot (r')^{-\rho} d\mu \ge \frac{\text{sgn}(\rho)}{\|q\|} \int p_{\lambda} \cdot (q')^{-\rho} d\mu \cdot \int q \cdot (r')^{-\rho} d\mu,
\end{eqnarray*}
i.e.,
\begin{eqnarray*}
  \mathscr{I}_{\alpha}(P_{\lambda},R)&\ge& \mathscr{I}_{\alpha}(P_{\lambda},Q)+\mathscr{I}_{\alpha}(Q,R)\\
    & \ge & \mathscr{I}_{\alpha}(Q,R).
\end{eqnarray*}
This completes the proof of (a).

(b) The ``if'' part is a trivial consequence of (a). We proceed to prove the ``only if'' part.

The finiteness of $\mathscr{I}_{\alpha}(Q,R)$ implies that $\mathscr{I}_{\alpha}(P,R)$ and $\mathscr{I}_{\alpha}(S,R)$ are also finite. Indeed, from (\ref{p1:eqn:convex combination}), it is clear that $p \le \lambda^{-1} q$ and thus $p/r \le \lambda^{-1} q/r$. As a consequence, we have
\begin{eqnarray*}
  \left(\frac{p'}{r'}\right)^{\frac{1}{\alpha}}
    & = & \frac{p}{r}\cdot \frac{\|r\|}{\|p\|} \\
    & \le & {\lambda}^{-1}\frac{q}{r}\cdot \frac{\|r\|}{\|p\|} \\
    & = & {\lambda}^{-1}\left(\frac{q'}{r'}\right)^{\frac{1}{\alpha}}\cdot \frac{\|q\|}{\|p\|}.
\end{eqnarray*}
Integrating with respect to $R'$, we get
\[
  \int \left(\frac{p'}{r'}\right)^{\frac{1}{\alpha}} dR'
  \le {\lambda}^{-1}\frac{\|q\|}{\|p\|}\cdot \int \left(\frac{q'}{r'}\right)^{\frac{1}{\alpha}}dR'
  < \infty.
\]
From (\ref{p1:eqn:alphadiv}), we have
\begin{equation*}
\I_{\alpha}(P,Q) = \frac{1}{\rho} \cdot \log \left[\text{sgn}(\rho) \int \left(\frac{p'}{q'}\right)^{\frac{1}{\alpha}} \, dQ' - 1\right].
\end{equation*}
Hence $\mathscr{I}_{\alpha}(P,R) \leq \mathscr{I}_{\alpha}(Q,R) + c$ for some constant $c$, and therefore $\mathscr{I}_{\alpha}(P,R)$ is finite. Similarly $\mathscr{I}_{\alpha}(S,R)$ is also finite.

Applying the first part of the theorem, we get
\begin{eqnarray*}
  \mathscr{I}_{\alpha}(P,R) & \ge & \mathscr{I}_{\alpha}(P,Q)+\mathscr{I}_{\alpha}(Q,R) \\
  \mathscr{I}_{\alpha}(S,R) & \ge & \mathscr{I}_{\alpha}(S,Q)+\mathscr{I}_{\alpha}(Q,R).
\end{eqnarray*}
The first inequality is the same as (\ref{p1:eqn:equivalentpythagoreanidentity}) while the second inequality is the same as (\ref{p1:eqn:equivalentpythagoreanidentity}) with $s$, the density of $S$, in place of $p$. Suppose one of these were a strict inequality. Then the $\lambda$ and $(1-\lambda)$ weighted linear combination of these two inequalities, along with $Q=\lambda P+(1-\lambda)S$, yields (\ref{p1:eqn:equality}) with a strict inequality, which is the same as (\ref{p1:eqn:trivial-statement}) with a strict inequality, a contradiction. So the two inequalities must be equalities. This proves the ``only if'' part and completes the proof of (b).
\end{IEEEproof}

\vspace{0.1in}

Once Theorem \ref{p1:thm:pythagorean} is established for general measure spaces, the proofs of the following results are exactly as in \cite{200701TIT_Sun}. We provide them for the benefit of the reader and for ease of reference. Let us first recall that any $Q \in \mathbb{E}$ is said to be an {\em algebraic inner point} of $\mathbb{E}$ if for every $P \in \mathbb{E}$ there exists $S \in \mathbb{E}$ and $0 < t < 1$ such that $Q = tP + (1-t)S$.

\vspace{0.1in}

\begin{figure}[tb]
 \centering
 \includegraphics[width=.2\linewidth]{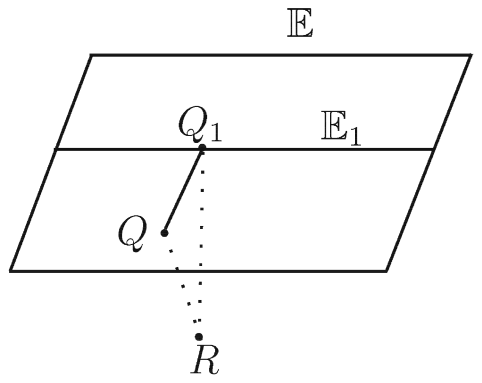}
 \caption{\label{p1:fig:transitivity}Subspace-transitivity}
\end{figure}

\begin{thm} The following statements hold.
\label{p1:thm:pyth_and_proj}
\begin{itemize}
\item[(a)] ({\em Projection and the Pythagorean property}): A probability measure $Q\in \mathbb{E} \cap B(R,\infty)$ is a forward $\mathscr{I}_{\alpha}$-projection of $R$ on the convex set $\mathbb{E}$ of probability measures if and only if every $P \in \mathbb{E} \cap B(R,\infty)$ satisfies (\ref{p1:eqn:pythagorean_inequality}). If the forward $\mathscr{I}_{\alpha}$-projection is an algebraic inner point of $\mathbb{E}$ then $\mathbb{E} \subset B(R,\infty)$ and (\ref{p1:eqn:pythagorean_equality}) holds for every $P \in \mathbb{E}$.

\item[(b)] ({\em Subspace-transitivity}): Let $\mathbb{E}$ and $\mathbb{E}_1 \subset \mathbb{E}$ be convex sets of probability measures. Let $R$ have the forward $\mathscr{I}_{\alpha}$-projection $Q$ on $\mathbb{E}$ and the forward $\mathscr{I}_{\alpha}$-projection $Q_1$ on $\mathbb{E}_1$, and suppose that (\ref{p1:eqn:pythagorean_equality}) holds for every $P \in \mathbb{E}$. Then $Q_1$ is the forward $\mathscr{I}_{\alpha}$-projection of $Q$ on $\mathbb{E}_1$. (See figure \ref{p1:fig:transitivity}).
\end{itemize}
\end{thm}

\vspace{0.1in}

\begin{remark}
 Th. \ref{p1:thm:pyth_and_proj} (a) essentially says that the forward projection is the unique point in $\mathbb{E}$ that has the Pythagorean property. The importance of Th. \ref{p1:thm:pyth_and_proj} (b) will be made clear in section \ref{p1:sec:transitivity}.
\end{remark}

\begin{IEEEproof}
(a) Consider the first part of the statement. The ``if'' part is trivial from the nonnegativity of $\mathscr{I}_{\alpha}$. The ``only if'' part easily follows from Theorem \ref{p1:thm:pythagorean}-(a). Indeed, $Q\in \mathbb{E}\cap B(R,\infty)$ is the forward $\mathscr{I}_{\alpha}$-projection of $R$ implies that for every $P\in \mathbb{E}$, we have $\mathscr{I}_{\alpha}(P_{\lambda},R)\ge \mathscr{I}_{\alpha}(Q,R)$ where $P_{\lambda} = \lambda P + (1-\lambda) Q$. Hence by Theorem \ref{p1:thm:pythagorean}-(a), (\ref{p1:eqn:pythagorean_inequality}) holds.

If the forward $\mathscr{I}_{\alpha}$-projection $Q$ is an algebraic inner point of $\mathbb{E}$ then for every $P\in \mathbb{E}$, there exists $S \in \mathbb{E}$ and $\lambda \in (0,1)$ such that $Q=\lambda P+(1-\lambda)S$. Hence by Theorem \ref{p1:thm:pythagorean}-(b),  (\ref{p1:eqn:pythagorean_equality}) holds.

(b) Applying Theorem \ref{p1:thm:pyth_and_proj}-(a) to $\mathbb{E}_1$, we have for every $P\in \mathbb{E}_1$
\begin{eqnarray*}
 \mathscr{I}_{\alpha}(P,R) &\ge& \mathscr{I}_{\alpha}(P,Q_1)+\mathscr{I}_{\alpha}(Q_1,R)\\
   &=& \mathscr{I}_{\alpha}(P,Q_1)+(\mathscr{I}_{\alpha}(Q_1,Q)+\mathscr{I}_{\alpha}(Q,R)),
\end{eqnarray*}
where the second equality follows from the equality hypothesis that (\ref{p1:eqn:pythagorean_equality}) holds. Using this same equality hypothesis, we also have
\[
  \mathscr{I}_{\alpha}(P,R)=\mathscr{I}_{\alpha}(P,Q)+\mathscr{I}_{\alpha}(Q,R).
\]
Thus
\[
  \mathscr{I}_{\alpha}(P,Q) \geq \mathscr{I}_{\alpha}(P,Q_1)+\mathscr{I}_{\alpha}(Q_1,Q)
\]
for every $P\in \mathbb{E}_1$. Applying Theorem \ref{p1:thm:pyth_and_proj}-(a) once again, we conclude that $Q_1$ is the forward $\mathscr{I}_{\alpha}$-projection of $Q$ on $\mathbb{E}_1$.
\end{IEEEproof}

\vspace*{.1in}

Theorem \ref{p1:thm:pyth_and_proj}-(a) yields a simple proof of the uniqueness of projection on a convex $\mathbb{E}$, if the projection exists. Indeed, let $Q_1$ and $Q_2$ be two projections of a probability measure $R$ on a convex $\mathbb{E}$. Then $\mathscr{I}_{\alpha}(Q_1,R)=\mathscr{I}_{\alpha}(Q_2,R)<\infty$. By Theorem \ref{p1:thm:pyth_and_proj}-(a),
\[
  \mathscr{I}_{\alpha}(Q_2,R)\ge \mathscr{I}_{\alpha}(Q_2,Q_1)+\mathscr{I}_{\alpha}(Q_1,R).
\]
Canceling $\mathscr{I}_{\alpha}(Q_2,R)$ and $\mathscr{I}_{\alpha}(Q_1,R)$, we get $\mathscr{I}_{\alpha}(Q_2,Q_1)=0$ which further implies $Q_1=Q_2$.

\section{Example: Forward $\mathscr{I}_{\alpha}$-projection for a linear family}
\label{p1:sec:projection_for_linear_family}
In this section we provide an explicit characterization of the forward $\mathscr{I}_{\alpha}$-projection on a linear family.

Let $\Gamma$ be an arbitrary index set and let $f_{\gamma}\colon\mathbb{X}\to \mathbb{R}$, for $\gamma \in \Gamma$, be measurable functions. The family of probability measures defined by
\begin{equation}
  \label{p1:linear_family}
  \mathbb{L}=\left\{ P \colon \int f_{\gamma} \,dP = 0, \gamma \in \Gamma \right\},
\end{equation}
if nonempty, is called a \emph{linear family}\footnote{Let us reiterate the standing assumptions: $P \ll \mu$ and the $\mu$-density $p \in L^{\alpha}(\mu)$ for every $P \in \mathbb{L}$.}.

Our next result is that the forward $\mathscr{I}_{\alpha}$-projection on a linear family is a member of an associated $\alpha$-{\em power-law family}\footnote{A parametric family of probability distributions that are of the form (\ref{p1:eqn:density_power_law1}).} just as forward $\mathscr{I}$-projection on a linear family is a member of an associated exponential family \cite[Th.~3.1]{1975xxAP_Csi}. The proof for $\alpha < 1$ is similar with only minor changes. The proof for $\alpha > 1$ involves some additional conditions. We will explore the geometric relationship between the linear family and the $\alpha$-power-law family in a companion paper \cite{2014xxManuscript2_KumSun}.

\vspace*{.1in}

\begin{thm}
  \label{p1:thrm:density}
  Let $\alpha > 0$ and $\alpha \neq 1$. Let $\mathbb{L}$ be a linear family of probability measures as in (\ref{p1:linear_family}). Let $R$ have $\mu$-density $r$.

  (a) If $Q$ is the forward $\mathscr{I}_{\alpha}$-projection of $R$ on $\mathbb{L}$ then the $\mu$-density $q$ of $Q$ satisfies
  \begin{align}
   \label{p1:eqn:density_power_law1}
    q(x)^{\alpha-1} & =  c \cdot r(x)^{\alpha-1} + g(x), & \forall x \notin N \\
   \label{p1:eqn:density_power_law2}
    q(x) & = 0, & \forall x \in N,
  \end{align}
  where $N\subseteq \X$ is such that, for every $P \in \mathbb{L} \cap B(R,\infty)$,
  \begin{eqnarray}
    \label{p1:eqn:density_power_law3}
    \left\{
    \begin{array}{rll}
     P(N)  & = \quad 0,
             & \mbox{ if } \alpha < 1 \\
     \displaystyle
     c \int_N r^{\alpha-1} ~dP & \leq \quad \displaystyle\int_{\mathbb{X} \setminus N} g ~dP,
            & \mbox{ if } \alpha > 1, \\
    \end{array}
    \right.
  \end{eqnarray}

  \begin{eqnarray}
    \label{p1:eqn:c=}
    c = \frac{\int q^{\alpha}\,d\mu}{\int q r^{\alpha-1}\,d\mu},
  \end{eqnarray}
  and $g$ belongs to the $L^1(Q)$-closure of the linear space spanned by $\{f_\gamma\}_{\gamma \in \Gamma}$.

  (b) Conversely, if there is a $Q\in \mathbb{L}$ whose $\mu$-density satisfies (\ref{p1:eqn:density_power_law1})-(\ref{p1:eqn:density_power_law3}) for some scalar $c$ and some $g$ in the linear span of $\{f_\gamma\}_{\gamma \in \Gamma}$, then $Q$ is the forward $\mathscr{I}_{\alpha}$-projection of $R$ on $\mathbb{L}$, (\ref{p1:eqn:pythagorean_inequality}) holds for every $P \in \mathbb{L} \cap B(R,\infty)$, and further, (\ref{p1:eqn:pythagorean_inequality}) holds with equality when $\alpha < 1$.
\end{thm}

\vspace*{.1in}

\begin{IEEEproof}
(a) Let $Q$ be the forward $\mathscr{I}_{\alpha}$-projection of $R$ on $\mathbb{L}$ with $\mu$-density $q$. Let $N = \{ x \in \mathbb{X} \colon q(x) = 0 \}$. By definition of the forward $\mathscr{I}_{\alpha}$-projection, we have $\mathscr{I}_{\alpha}(Q,R) < \infty$. When $\alpha < 1$, if $P \in \mathbb{L} \cap B(R, \infty)$, then Theorem \ref{p1:thm:pyth_and_proj}-(a) implies (\ref{p1:eqn:pythagorean_inequality}), which further implies $\mathscr{I}_{\alpha}(P,Q) < \infty$, $P \ll Q$, and thus $P(N) = 0$. We will soon define $g$ on $\mathbb{X}\setminus N$ and will show the inequality in (\ref{p1:eqn:density_power_law3}) for $\alpha > 1$ later in this proof.

From $\mathscr{I}_{\alpha}(Q,R) < \infty$, using (\ref{p1:alphadiv_expanded_general}), it is also easy to verify that $0 < \int q \, r^{\alpha -1} \, d\mu < \infty$. Define
\[
  \mathbb{L}_1 := \{P\in \mathbb{L}\colon p(x)\le 2 q(x)\,\text{a.e.}[\mu]\}.
\]
Obviously, $\mathbb{L}_1$ is convex and $Q \in \mathbb{L}_1$. For any $P \in \mathbb{L}_1$, define $P_1$ to have the density $p_1(x)=2q(x)-p(x)$. We then have $P_1 \in \mathbb{L}_1$ and $Q = \frac{P+P_1}{2}$. Hence $Q$ is an algebraic inner point of $\mathbb{L}_1$. By Theorem \ref{p1:thm:pyth_and_proj}-(a), (\ref{p1:eqn:pythagorean_inequality}) holds with equality for all $P \in \mathbb{L}_1$. This equality can be simplified, based on (\ref{p1:alphadiv_linear_form_general}), to
\begin{eqnarray}
  \label{p1:eqn:pyth_equal_simp}
  \int p r^{\alpha-1} d\mu & = & \int p q^{\alpha-1} d\mu\cdot \frac{\int q r^{\alpha-1} d\mu}{\int q^{\alpha} d\mu}\\
   & = & c^{-1} \int p q^{\alpha-1} d\mu,
\end{eqnarray}
where $c$ is given by (\ref{p1:eqn:c=}). This can be rewritten as
\begin{eqnarray}
  \int p \cdot \left(q^{\alpha-1}-c r^{\alpha-1}\right) \,d\mu = 0 \quad \forall P \in \mathbb{L}_1,
\end{eqnarray}
which with $g(x) := q(x)^{\alpha-1} - c r(x)^{\alpha-1}, ~ x \in \mathbb{X} \setminus N,$ is the same as
\begin{eqnarray}
  \label{p1:eqn:pyth_equal_simp1}
  \int p g \, d\mu = 0 \quad \forall P \in \mathbb{L}_1.
\end{eqnarray}

We have left $g$ undefined for $x$ with $q(x) = 0$, but this is inconsequential because we now show $g$ belongs to the $L^1(Q)$-closure of the linear span of $\{f_\gamma\}_{\gamma \in \Gamma}$.

From (\ref{p1:eqn:pyth_equal_simp1}), we get
\begin{eqnarray}
  \label{p1:eqn:pyth_equal_simp1a}
  \int  g \cdot \frac{dP}{dQ} \cdot \, dQ = 0 \quad \forall P \in \mathbb{L}_1,
\end{eqnarray}
and by setting $P = Q$ in (\ref{p1:eqn:pyth_equal_simp1a}) we get
\begin{eqnarray}
  \label{p1:eqn:pyth_equal_simp2}
  \int g \, dQ = 0.
\end{eqnarray}
Combining (\ref{p1:eqn:pyth_equal_simp1a}) and (\ref{p1:eqn:pyth_equal_simp2}) yields
\begin{eqnarray}
  \label{p1:eqn:pyth_equal_change_of_measure}
  \int g \left(\frac{dP}{dQ}-1\right) \,dQ = 0 \quad \forall P \in \mathbb{L}_1.
\end{eqnarray}

If $h \colon\mathbb{X} \to \mathbb{R}$ is a measurable function such that $|h|\le 1,\, \text{a.e.}[Q]$, and further
\begin{eqnarray}
  \label{p1:functional_being_zero}
  \int h \,dQ=0, \text{ and } \int f_\gamma h \,dQ =0 \text{ for every } \gamma \in \Gamma,
\end{eqnarray}
then $P$ defined according to $dP=(h+1)\,dQ$ belongs to $\mathbb{L}_1$, and from (\ref{p1:eqn:pyth_equal_change_of_measure}), it follows that
\begin{eqnarray}
  \label{p1:eqn:ghdQ=0}
  \int g h \,dQ = 0.
\end{eqnarray}
It immediately follows after scaling that if $h\in L^{\infty}(Q)$, the dual of $L^1(Q)$, and (\ref{p1:functional_being_zero}) holds, then (\ref{p1:eqn:ghdQ=0}) must also hold. In other words, any continuous linear functional $F_h\colon L^1(Q) \rightarrow \mathbb{R}$ given by $F_h(f) = \int f h \, dQ$ that vanishes on the linear subspace spanned by $1$ and the $f_{\gamma}$'s also vanishes at $f=g$. By the Hahn-Banach theorem \cite[Th.~5.8.a]{1999xxRA_Fol}, $g$ is in the $L^1(Q)$-closure of that linear subspace. From (\ref{p1:eqn:pyth_equal_simp2}), it follows that $g$ is in the $L^1(Q)$-closure of the subspace spanned by the $f_{\gamma}$'s alone.

We now show the inequality in (\ref{p1:eqn:density_power_law3}) for $\alpha > 1$. For any $P \in \mathbb{L} \cap B(R,\infty)$, where such a $P$ may be outside $\mathbb{L}_1$, let us observe that
\begin{eqnarray}
  \displaystyle
  \label{p1:eqn:pyth-app}
  0 & \leq & \int_{\X} p q^{\alpha-1} d\mu - c \int_{\X} p r^{\alpha-1} d\mu\\
    \displaystyle
    \label{p1:eqn:pyth-restrict}
    & = & \int_{\mathbb{X} \setminus N} p q^{\alpha-1} d\mu - c \int_{\X} p r^{\alpha-1} d\mu \\
    \displaystyle
    \label{p1:eqn:pyth-subst}
    & = & \int_{\mathbb{X} \setminus N} p \cdot ( c r^{\alpha-1} + g) ~ d\mu ~ - ~ c \int_{\X} p r^{\alpha-1} d\mu \\
    \displaystyle
    \label{p1:eqn:pyth-cond}
    & = & \int_{\mathbb{X} \setminus N} p g ~ d\mu ~ - ~ c \int_N p r^{\alpha-1} d\mu,
\end{eqnarray}
where (\ref{p1:eqn:pyth-app}) follows from the combination of (\ref{p1:alphadiv_expanded_general}), (\ref{p1:eqn:pythagorean_inequality}), and (\ref{p1:eqn:c=}); consequently, (\ref{p1:eqn:pyth-restrict}) follows from the fact that $q(x) = 0$ for $x \in N$, (\ref{p1:eqn:pyth-subst}) follows from the definition of $g(x)$ on the set $x \in \mathbb{X} \setminus N$, and (\ref{p1:eqn:pyth-cond}) follows from the cancellation of a portion of the last integral term on the right-hand side of (\ref{p1:eqn:pyth-subst}). Inequality (\ref{p1:eqn:density_power_law3}) for $\alpha > 1$ follows from (\ref{p1:eqn:pyth-cond}). This completes the proof of (a).

(b) Let $Q\in \mathbb{L}$ have $\mu$-density $q$ which satisfies (\ref{p1:eqn:density_power_law1})-(\ref{p1:eqn:density_power_law3}) where $c$ is some scalar and $g$ is a linear combination of the $f_{\gamma}$'s; so $\int g \, dP = 0$ for all $P \in \mathbb{L}$. Integrating (\ref{p1:eqn:density_power_law1})-(\ref{p1:eqn:density_power_law2}) with respect to $Q$ and using $\int g \, dQ = 0$, we get
 \[
   \int q^{\alpha} \, d\mu = c \int q r^{\alpha-1} \, d\mu
 \]
 from which the following are clear:
 \begin{itemize}
   \item $0 < \int q r^{\alpha-1} \, d\mu < \infty$, and so $\mathscr{I}_{\alpha}(Q,R) < \infty$;
   \item $c > 0$ and satisfies (\ref{p1:eqn:c=}).
 \end{itemize}
 Fix any $P \in \mathbb{L}$ with $\mathscr{I}_{\alpha}(P,R) < \infty$. As claimed at the beginning of the proof of part (a), we then have $0 < \int p r^{\alpha-1} \, d\mu < \infty$. Integrating (\ref{p1:eqn:density_power_law1})-(\ref{p1:eqn:density_power_law2}) with respect to $P$, we now get
 \[
   \int p q^{\alpha-1} \, d\mu \geq c \int p r^{\alpha-1} \, d\mu,
 \]
 where
 \begin{itemize}
   \item equality holds when $\alpha < 1$ because of the assumption $P(N) = 0$,
   \item inequality holds when $\alpha > 1$ because of the inequality assumption in (\ref{p1:eqn:density_power_law3}); indeed, this assumption is the same as saying that the right-hand side of (\ref{p1:eqn:pyth-cond}) is $\geq 0$, and one proceeds in the reverse direction in that sequence of equalities to obtain the inequality (\ref{p1:eqn:pyth-app}) which is the same as the above inequality.
 \end{itemize}
 Since $c$ satisfies (\ref{p1:eqn:c=}), we have that (\ref{p1:eqn:pythagorean_inequality}) holds (with equality when $\alpha < 1$). By Theorem \ref{p1:thm:pyth_and_proj}-(a) (in the ``if'' direction) $Q$ is the forward $\mathscr{I}_{\alpha}$-projection of $R$ on $\mathbb{L}$.
\end{IEEEproof}

\vspace*{.1in}

\begin{remark}
\label{p1:remark:countereg}
As in the case of relative entropy ($\alpha = 1$), in Theorem \ref{p1:thrm:density}-(a), it is possible that the inequality in (\ref{p1:eqn:pythagorean_inequality}) is strict for some $P$ in the linear family, and in Csisz\'ar's words \cite[p.152]{1975xxAP_Csi}, ``neither the necessary nor the sufficient condition of Theorem \ref{p1:thrm:density} is both necessary and sufficient, in general.'' Csisz\'ar's counterexamples in \cite[pp.152-153]{1975xxAP_Csi}, but with $q^{\alpha-1} = c \cdot r^{\alpha-1} +g$ instead of $q = c \cdot r \cdot \exp \{ g \}$, continue to serve as counterexamples for our parametric setting (see Appendix \ref{p1:sec:app-countereg}).
\end{remark}

However, under an additional assumption, Theorem \ref{p1:thrm:density} can be leveraged to provide a necessary and sufficient condition for a $Q \in \mathbb{L}$ to be the forward $\mathscr{I}_{\alpha}$-projection.

\vspace*{.1in}

\begin{corollary}
\label{p1:cor:linear-family-iff}
Let $\alpha > 0$ and $\alpha \neq 1$. Let $\mathbb{L}$ be the linear family as defined in (\ref{p1:linear_family}). Suppose that the linear space spanned by $\{f_\gamma\}_{\gamma \in \Gamma}$ is $L^1(P)$-closed for every $P \in \mathbb{L}$. Consider a $Q \in \mathbb{L}$. $Q$ is the forward $\mathscr{I}_{\alpha}$-projection of $R$ on $\mathbb{L}$ if and only if the $\mu$-density $q$ of $Q$ satisfies (\ref{p1:eqn:density_power_law1})-(\ref{p1:eqn:density_power_law3}) for some scalar $c$ and some $g$ in the span of $\{f_\gamma\}_{\gamma \in \Gamma}$. Moreover, the inequality in (\ref{p1:eqn:density_power_law3}) for $\alpha > 1$ is equivalent to
\begin{equation}
  \label{p1:eqn:density_power_law4}
  \int_N (c r^{\alpha-1} + g) ~dP \leq 0, \quad \alpha > 1.
\end{equation}
\end{corollary}

\vspace*{.1in}

\begin{IEEEproof}
The forward direction is immediate from the forward direction of Theorem \ref{p1:thrm:density} and the hypothesis that the linear space spanned by $\{f_\gamma\}_{\gamma \in \Gamma}$ is $L^1(Q)$-closed; so $g$ is in the span of $\{f_\gamma\}_{\gamma \in \Gamma}$. The reverse direction is the same as the reverse direction in Theorem \ref{p1:thrm:density}.

To prove (\ref{p1:eqn:density_power_law4}), let us observe that because $g$ is in the span of $\{f_\gamma\}_{\gamma \in \Gamma}$, it is well-defined for all $x \in \mathbb{X}$ and consequently satisfies $\int g ~dP = 0$ for every $P \in \mathbb{L}$. Adding $\int_N g ~dP$ to both sides of (\ref{p1:eqn:density_power_law3}) and using $\int g~dP = 0$, we get (\ref{p1:eqn:density_power_law4}).
\end{IEEEproof}

\vspace*{.1in}

One example where the linear space spanned by $\{f_\gamma\}_{\gamma \in \Gamma}$ is $L^1(P)$-closed for every $P \in \mathbb{L}$ is when $\Gamma$ is finite, i.e., $\Gamma = \{ 1, 2, \ldots, k\}$ for some finite $k$. If $Q$ is the forward $\mathscr{I}_{\alpha}$-projection of $R$ on $\mathbb{L}$, then the expression
\[
  q(x)^{\alpha-1} = c \cdot r(x)^{\alpha-1} + \sum_{\gamma=1}^k \theta_\gamma f_{\gamma}(x),
\]
where $(\theta_1, \ldots, \theta_k) \in \mathbb{R}^k$, holds for all $x$ with $q(x) > 0$. Moreover, (\ref{p1:eqn:pythagorean_inequality}) holds for all $P \in \mathbb{L} \cap B(R,\infty)$, and it holds with equality when $\alpha < 1$.

For relative entropy, $\alpha = 1$, Csisz\'ar provides another example: the family of probability measures on a product space $\mathbb{X} = \mathbb{X}_1 \times \mathbb{X}_2$ with the associated product $\sigma$-algebra, having specified marginals. We leave the question of whether Corollary \ref{p1:cor:linear-family-iff} is applicable or not to this setting as an open question.

Even though Corollary \ref{p1:cor:linear-family-iff} characterizes the forward $\mathscr{I}_{\alpha}$-projection to some extent, existence of the projection is not assured, and one appeals to Theorem \ref{p1:thm:projection} or other means to guarantee existence. Let us note in passing two instances when the crucial hypothesis of Theorem \ref{p1:thm:projection}, that the set of $\mu$-densities is $L^{\alpha}(\mu)$-closed, holds.
\begin{itemize}
  \item[(a)] If $\alpha > 1$, $\mu(\mathbb{X}) < +\infty$, and $f_{\gamma} \in L^{\infty}(\mu)$ for $\gamma = 1, \ldots, k$, then a simple application of Lyapunov's inequality\footnote{Lyapunov's moment inequality states that if $\mu(\X) <\infty$, $0 < r < s \le \infty$,  and $u\in L^s(\mu)$, then $\|u\|_r\le \|u\|_s {\mu(\X)}^{(\nicefrac{1}{r}) - (\nicefrac{1}{s})}$, and consequently $L^s(\mu)\subseteq L^r(\mu)$.} and the dominated convergence theorem suffices to show that $\mathcal{L}$, the set of $\mu$-densities of probability measures in $\mathbb{L}$, is $L^{\alpha}(\mu)$-closed.
  \item[(b)] If $\mathbb{X}$ is finite, point-wise convergence suffices to establish that $\mathcal{L}$ is $L^{\alpha}(\mu)$-closed.
\end{itemize}

\vspace*{.1in}

Let us now exploit the understanding we have gained to generalize Lemma \ref{p1:lemma:properties}-e) on R\'enyi entropy maximizers.

\vspace*{.1in}

\begin{corollary}
\label{p1:cor:renyi-maximizer}
Let $\alpha > 0$ and $\alpha \neq 1$. Let $\mathbb{L}$ be the linear family as defined in (\ref{p1:linear_family}). If $\mathbb{L}$ has a member $Q$ whose $\mu$-density $q$ satisfies (\ref{p1:eqn:density_power_law1})-(\ref{p1:eqn:density_power_law3}) for some scalar $c$, some $g$ in the span of $\{f_\gamma\}_{\gamma \in \Gamma}$, and with $r(x) \equiv 1$, then
\begin{equation}
  \label{p1:eqn:renyi_maximizer}
  \mathscr{I}_{\alpha}(P, Q) \leq H_{\alpha}(Q) - H_{\alpha}(P) \quad \forall P \in \mathbb{L},
\end{equation}
with equality when $\alpha < 1$. Furthermore, $Q$ is the R\'{e}nyi entropy maximizer in $\mathbb{L}$.
\end{corollary}

\vspace*{.1in}

\begin{IEEEproof}
  It suffices to prove (\ref{p1:eqn:renyi_maximizer}). The second statement immediately follows.

  Using (\ref{p1:alphadiv_expanded_general}), (\ref{p1:eqn:renyi-entropy}), and after a simple rearrangement, we get
  \begin{eqnarray*}
    \mathscr{I}_{\alpha}(P,Q) & = & \frac{\alpha}{1-\alpha} \left[ \log \int p q^{\alpha-1}  d\mu - \log \int q^{\alpha} d\mu \right] + H_{\alpha}(Q) - H_{\alpha}(P).
  \end{eqnarray*}
  Let us note from (\ref{p1:eqn:density_power_law1})-(\ref{p1:eqn:density_power_law2}) that $\int q^{\alpha} d\mu = c$. So (\ref{p1:eqn:renyi_maximizer}) will hold if we can establish
  \begin{eqnarray*}
    \int p q^{\alpha-1} d\mu & = & c \quad \forall P \in \mathbb{L}, \quad \mbox{ if } \alpha < 1, \\
    \int p q^{\alpha-1} d\mu & \geq & c \quad \forall P \in \mathbb{L}, \quad \mbox{ if } \alpha > 1.
  \end{eqnarray*}
  Both of these are obvious from the hypotheses of the corollary via (\ref{p1:eqn:density_power_law1})-(\ref{p1:eqn:density_power_law3}), the assumption that $r(x) \equiv 1$, and the fact that $P$ and $Q$ are both probability measures belonging to $\mathbb{L}$.
\end{IEEEproof}

\vspace*{.1in}

\begin{remark}
When $0 < \mu(\mathbb{X}) < \infty$, with $r(x) \equiv 1$, define the probability measure $\tilde{R}$ with $\mu$-density 
\[
 \tilde{r}(x) := \frac{r(x)}{\mu(\mathbb{X})} \equiv \frac{1}{\mu(\mathbb{X})}.
\]
We then have from (\ref{p1:alphadiv_expanded_general}) that $\mathscr{I}_{\alpha}(P,\tilde{R}) = H_{\alpha}(\tilde{R}) - H_{\alpha}(P) = \log \mu(\mathbb{X}) - H_{\alpha}(P)$, and so the R\'enyi entropy maximizer on $\mathbb{L}$ is the forward $\mathscr{I}_{\alpha}$-projection of $\tilde{R}$ on $\mathbb{L}$. From (\ref{p1:alphadiv_linear_form_general}), it is clear that scale factors are irrelevant, and if we allow the second argument of $\mathscr{I}_{\alpha}$ to be positive measures, not just probability measures, then the R\'enyi entropy maximizer on $\mathbb{L}$ can be interpreted as the ``forward $\mathscr{I}_{\alpha}$-projection of $\mu$ on $\mathbb{L}$''. When $\mu(\mathbb{X})$ is not finite, there is no probability measure on $\mathbb{X}$ with the uniform $\mu$-density.  Nevertheless, Corollary \ref{p1:cor:renyi-maximizer} shows that the R\'enyi entropy maximizer is the ``forward $\mathscr{I}_{\alpha}$-projection of $\mu$ on $\mathbb{L}$''.
\end{remark}

\vspace*{.1in}

\begin{remark}
Student-t and Student-r distributions are maximizers of R\'enyi entropy under a covariance constraint \cite{200705AIHP_JohVig}. Since a Student-r distribution has a compact support, it can be shown to be the forward $\I_{\alpha}$-projection of the uniform distribution as described above, when $\alpha >1$. The support of a Student-t distribution is the whole of $\mathbb{R}^d$. However, it can also be seen as a limit of forward $\I_{\alpha}$-projections of uniform distributions on an increasing sequence of compact subsets of $\mathbb{R}^d$, when $\alpha <1$.
\end{remark}

\section{Transitivity and Iterated Projections for a linear family}
\label{p1:sec:transitivity}

In this section we assume $\mathbb{X}$ is finite. Let $\mathcal{P}(\mathbb{X})$ be the space of all probability measures on $\mathbb{X}$. In a remarkable paper \cite{1991xxTAS_Csi} on an axiomatic approach to inference, Csisz\'ar explored some natural axioms for {\em selection} and {\em projection rules}, and their consequences on linear families.

A {\em projection rule} is a mapping that (in our context) takes a probability measure $R$ and a linear family $\mathbb{L}$ and maps them to a probability measure $\Pi(\mathbb{L}|R)$ in $\mathbb{L}$, such that if $R \in \mathbb{L}$ then $\Pi(\mathbb{L}|R) = R$. $\Pi(\mathbb{L}|R)$ is then called the projection of $R$ on $\mathbb{L}$. A projection rule is said to be generated by a function $F(P|R), ~P \in \mathcal{P}(\mathbb{X}), ~R \in \mathcal{P}(\mathbb{X})$, if for each $R$, $\Pi(\mathbb{L}|R)$ is the unique element of $\mathbb{L}$ where $F(P|R)$ is minimized subject to $P \in \mathbb{L}$. A projection rule may be interpreted as follows: a ``prior guess'' $R$ is updated to $\Pi(\mathbb{L}|R)$ upon information that the ``feasible set'' is $\mathbb{L}$.

Clearly, the forward $\mathscr{I}_{\alpha}$-projection of $R$ on a linear family $\mathbb{L}$ is an example of a projection rule that is generated by the function $F(P|R) = \mathscr{I}_{\alpha}(P, R)$. Csisz\'ar \cite[Th.~1]{1991xxTAS_Csi} showed that any {\em regular} and {\em local} projection rule, see \cite[Def.~2-3]{1991xxTAS_Csi} for the definitions, is generated by a separable function $F(P|R) = \sum_{x \in \mathbb{X}} \phi_x(P(x)|R(x))$, for some component functions $\phi_x(\cdot | \cdot), x \in \mathbb{X}$, with the value 0 at $P = R$.

Another desired property of a projection rule is subspace-transitivity (\cite[Def.~6]{1991xxTAS_Csi}). A projection rule is {\em subspace-transitive} if for any $\mathbb{L}' \subset \mathbb{L}$, both of which are linear families, and any probability measure $R$, we have
\[
  \Pi(\mathbb{L}'|R) = \Pi(\mathbb{L}'| \Pi(\mathbb{L}| R)).
\]
This can be interpreted as follows: if a ``prior guess'' $R$ is updated to $\Pi(\mathbb{L}|R)$ upon information that the ``feasible set'' is $\mathbb{L}$, and further information restricts the possibilities to a smaller feasible set $\mathbb{L}'$, then updating the ``current guess'' $\Pi(\mathbb{L}|R)$ on the basis of all available information yields the same outcome as updating the ``prior guess'' $R$ directly on the basis of all available information. Csisz\'ar showed \cite[Th.~3]{1991xxTAS_Csi} that any regular, local, and subspace-transitive projection rule is generated by Bregman's divergence of the sum-form, i.e.,
\[
  F(P|R) = \Phi(P) - \Phi(R) - \langle \mbox{grad } \Phi(R), P - R \rangle,
\]
where $\Phi(P) = \sum_x \varphi_x(P(x))$. Squared Euclidean distance and relative entropy $\mathscr{I}_1$ are examples of such divergences.

$\mathscr{I}_{\alpha}$ is, in general, neither of the sum-form nor a Bregman's divergence. Yet when $\alpha < 1$, the projection rule generated by $\mathscr{I}_{\alpha}(P,R)$ is subspace-transitive. The property fails in general when $\alpha > 1$, but holds even in this case in the special circumstance when the projection is an algebraic inner point. The main goal of this section is to establish subspace transitivity. This suggests that if one is willing to forgo the locality axiom of a projection rule, then there is at least one other family of projection rules, those generated by $\mathscr{I}_{\alpha}$, that are regular and subspace-transitive.

To formalize the result, we begin with two simple propositions. For a probability measure $P$ write $\text{Supp}(P)$ for the set of $x$ where $P(x) > 0$. For a family of probability measures $\mathbb{E}$, write $\text{Supp}(\mathbb{E})$ for the union of the supports of all probability measures in $\mathbb{E}$. We then have the following.

\vspace*{.1in}

\begin{proposition}
\label{p1:prop:fullsupp}
Let $\alpha < 1$. Let $Q$ be the forward $\mathscr{I}_{\alpha}$-projection of $R$ on $\mathbb{E}$. If $\mathbb{E}$ is convex, then $\text{Supp}(Q) = \text{Supp}(\mathbb{E})\cap \text{Supp}(R)$.
\end{proposition}

\vspace*{.1in}

\begin{IEEEproof}
We may restrict attention to those $P\in \mathbb{E}$ such that $P\ll R$. For such a $P$, let $P_t = (1-t)Q+tP$, $0 \le t \le 1$ . Since $\mathbb{E}$ is convex, $P\in \mathbb{E}$ implies that $P_{t}\in \mathbb{E}$. By the mean value theorem, for each $t \in (0,1)$, there exists $\tilde{t}\in (0,t)$ such that
\begin{eqnarray}
\label{p1:derivative1}
0\le \frac{1}{t}\Big[\mathscr{I}_{\alpha}(P_{t},R)-\mathscr{I}_{\alpha}(Q,R)\Big] = \frac{d}{ds}\mathscr{I}_{\alpha}(P_{s},R)|_{s=\tilde{t}}.
\end{eqnarray}
The first inequality follows from the fact that $Q$ is the projection. Using (\ref{p1:alphadiv_expanded_discrete}), we see that
\begin{eqnarray}
\label{p1:derivative2}
\frac{d}{ds}\mathscr{I}_{\alpha}(P_{s},R) = \frac{\alpha}{1-\alpha} \left[ \frac{\sum_x(P(x)-Q(x))R(x)^{\alpha-1}}{\sum_x P_{s}(x)R(x)^{\alpha-1}} - \frac{\sum_x(P(x)-Q(x))P_{s}(x)^{\alpha-1}}{\sum_x P_{s}(x)^{\alpha}} \right].
\end{eqnarray}
Suppose $Q(x)=0$ for an $x \in \text{Supp}(P)$. Then $\alpha<1$ implies that right-hand side of (\ref{p1:derivative2}) goes to $-\infty$ as $t\downarrow 0$, which contradicts the nonnegativity requirement in (\ref{p1:derivative1}). Hence $\text{Supp}(P)\subset \text{Supp}(Q)$ for every $P\in \mathbb{L}$. Also, since $Q$ is the $\mathscr{I}_{\alpha}$-projection of $R$, $\mathscr{I}_{\alpha}(Q,R) <\infty$, and as a consequence, $\text{Supp}(Q) \subset \text{Supp}(R)$. This establishes the proposition.
\end{IEEEproof}

\vspace*{.1in}

Consider now the linear family of probability measures on $\mathbb{X}$ given by
\begin{eqnarray}
\label{p1:linear_family_discrete}
 \mathbb{L} = \Big\{P\colon \sum_x P(x)f_{\gamma}(x) = 0, \text{ for all }\gamma = 1,\dots,k\Big\}.
\end{eqnarray}
Since $\mathbb{X}$ is finite, we already saw at the end of the previous section that $\mathbb{L}$ is closed in $L^{\alpha}(\mu)$, with $\mu$ being the counting measure. By Theorem \ref{p1:thm:projection}, any probability measure $R$ with $\mathscr{I}_{\alpha}(P,R) < \infty$ for some $P \in \mathbb{L}$ has a forward $\mathscr{I}_{\alpha}$-projection on $\mathbb{L}$. Moreover, we have the following.

\vspace*{.1in}

\begin{proposition}
\label{p1:prop:alg_inner_point}
Let $\alpha < 1$. Let $R$ have full support. Let $\mathbb{L}$ be as in (\ref{p1:linear_family_discrete}) and let $Q$ be the forward $\mathscr{I}_{\alpha}$-projection of $R$ on $\mathbb{L}$. Then $Q$ is an algebraic inner point of $\mathbb{L}$.
\end{proposition}

\vspace*{.1in}

\begin{IEEEproof}
By Proposition \ref{p1:prop:fullsupp}, $\text{Supp}(Q)=\text{Supp}(\mathbb{L})$. Hence for every $P\in \mathbb{L}$, one can find $t<0$ such that $P_t=(1-t)Q+tP\in \mathbb{L}$. This implies that
\[
  Q=\frac{1}{1-t}P_t-\frac{t}{1-t}P,
\]
and hence $Q$ is an algebraic inner point of $\mathbb{L}$.
\end{IEEEproof}

\vspace*{.1in}

We are now ready to state the main result of this section.

\vspace*{.1in}

\begin{theorem}[Subspace-transitivity]
\label{p1:thm:transitivity}
Let $\mathbb{L}_1 \subset \mathbb{L}$ be two linear families of probability measures.  Let $R$ be a probability measure with full support. Let $R$ have the forward $\mathscr{I}_{\alpha}$-projection $Q$ on $\mathbb{L}$ and the forward $\mathscr{I}_{\alpha}$-projection $Q_1$ on $\mathbb{L}_1$. If either (a) $\alpha < 1$ or (b) $\alpha > 1$ and $Q$ is an algebraic inner point of $\mathbb{L}$, then $Q_1$ is the forward $\mathscr{I}_{\alpha}$-projection of $Q$ on $\mathbb{L}_1$.
\end{theorem}

\vspace*{.1in}

\begin{IEEEproof}
If $\alpha < 1$, then by Proposition \ref{p1:prop:alg_inner_point}, $Q$ is an algebraic inner point of $\mathbb{L}$. If $\alpha > 1$, by assumption (b), $Q$ is an algebraic inner point of $\mathbb{L}$. Apply Theorem \ref{p1:thm:pyth_and_proj}-(a) to get that (\ref{p1:eqn:pythagorean_equality}) holds for all $P \in \mathbb{L}$. Now apply Theorem \ref{p1:thm:pyth_and_proj}-(b) to conclude that subspace-transitivity holds.
\end{IEEEproof}

\vspace*{.1in}

\begin{remark}
As can be observed from Theorem \ref{p1:thm:pyth_and_proj}-(b), and from the proof above, subspace-transitivity follows whenever there is equality in (\ref{p1:eqn:pythagorean_equality}). What is special about linear spaces under $\alpha < 1$ is that this equality comes for free, thanks to Proposition \ref{p1:prop:alg_inner_point}.
\end{remark}

\vspace*{.1in}

\begin{example}
The following example shows that subspace-transitivity for the $\mathscr{I}_{\alpha}$-projection rule need not hold when $\alpha>1$. Take $\alpha = 2$ and $\mathbb{X} = \{ 1,2,3,4 \}$. Take $R=(\nicefrac{1}{4},\nicefrac{1}{4},\nicefrac{1}{4},\nicefrac{1}{4})$. Consider the two linear families on the probability simplex in $\mathbb{R}^4$,
\begin{eqnarray*}
\mathbb{L} & = & \{P \in \mathcal{P}(\mathbb{X})\colon 8p_1 + 4p_2 + 2p_3 + p_4 = 7\}, \\
\mathbb{L}_1 & = & \{P \in \mathcal{P}(\mathbb{X})\colon 8p_1 + 4p_2 + 2p_3 + p_4 = 7;\, p_2 = \nicefrac{1}{8}\}.
\end{eqnarray*}
Thus
\[
 \mathbb{L} = \Big\{P \in \mathcal{P}(\mathbb{X})\colon \sum\limits_x P(x)f_1(x) = 0 \Big\},
\]
\[
 \mathbb{L}_1 = \Big\{P \in \mathcal{P}(\mathbb{X})\colon \sum\limits_x P(x)f_i(x) = 0, \, \, i=1,2 \Big\},
\]
where $f_1(\cdot) = (1,-3,-5,-6)$ and $f_2(\cdot) = (\nicefrac{-1}{8},\nicefrac{7}{8}, \nicefrac{-1}{8}, \nicefrac{-1}{8})$.

We claim that the forward $\mathscr{I}_{\alpha}$-projection of $R$ on $\mathbb{L}$ is $Q = (\nicefrac{3}{4}, \nicefrac{1}{4}, 0, 0)$. To check this claim, first note that $Q\in \mathbb{L}$. Also, with $c = \nicefrac{5}{2}$ and $\theta_1 = \nicefrac{1}{8}$, we can check that
\begin{eqnarray*}
  0 & < & Q(x) = c\, R(x) + \theta_1 f_1(x),\, x=1,2,\\
  0 & = & Q(3) = c\, R(3) + \theta_1 f_1(3),\\
  0 & = & Q(4) > c\, R(4) + \theta_1 f_1(4).  
\end{eqnarray*}
One can then easily verify that this $Q$ satisfies (\ref{p1:eqn:density_power_law4}) (which is equivalent to (\ref{p1:eqn:density_power_law3}) with $\alpha >1$) for every $P\in \mathbb{L}$. Hence, by Corollary \ref{p1:cor:linear-family-iff}, $Q$ is the forward $\mathscr{I}_{\alpha}$-projection of $R$ on $\mathbb{L}$.

Similarly one can show that the forward $\mathscr{I}_{\alpha}$-projection of $R$ on $\mathbb{L}_1$ is $Q_1 = (\nicefrac{19}{24}, \nicefrac{1}{8},\nicefrac{1}{12},0)$. Indeed, with $\theta_1 = \nicefrac{17}{144}$, $\theta_2 = \nicefrac{-7}{36}$ and $c = \nicefrac{187}{72}$, we have
\begin{eqnarray*}
  0 & < & Q_1(x) = c\, R(x) + \theta_1 f_1(x) +\theta_2 f_2(x),\, x=1,2,3,\\
  0 & = & Q_1(4) > c\, R(4) + \theta_1 f_1(4) +\theta_2 f_2(4).
\end{eqnarray*}
Again, $Q_1$ satisfies (\ref{p1:eqn:density_power_law4}) for every $P\in \mathbb{L}_1$ and, by Corollary \ref{p1:cor:linear-family-iff}, must be the forward $\mathscr{I}_{\alpha}$-projection of $R$ on $\mathbb{L}_1$.

Numerical calculations show that $(0.798, 0.125, 0.038, 0.039)$ is in $\mathbb{L}_1$ and
\[
 \mathscr{I}_{\alpha}((0.798,0.125,0.038,0.039), Q) = 0.0323 < \mathscr{I}_{\alpha}(Q_1,Q) = 0.0382.
\]
 If $\tilde Q_1$ is the forward $\mathscr{I}_{\alpha}$-projection of $Q$ on $\mathbb{L}_1$, it must satisfy $\mathscr{I}_{\alpha}(\tilde Q_1,Q) \le 0.0323$, which $Q_1$ does not. Thus, the transitive projection of $R$ on $\mathbb{L}_1$ via $Q$ is different from $Q_1$.
\end{example}

\vspace*{.1in}

The next theorem provides an iterative way of finding the forward $\mathscr{I}_{\alpha}$-projection for $\alpha <1$ when the set $\mathbb{L}$ is an  intersection of several linear families. A similar result is known for relative entropy ($\alpha=1$); see \cite[Th.~3.2]{1975xxAP_Csi}.

\vspace*{.1in}

\begin{theorem}[Iterated projections]
\label{p1:iterated_projections}
Let $\alpha < 1$. Suppose that $\mathbb{L}_0,\dots, \mathbb{L}_{m-1}$ are linear families of probability measures on a finite set $\mathbb{X}$ and that $\mathbb{L} = \bigcap_{i=0}^{m-1} \mathbb{L}_i \neq \emptyset$. Let $R$ be a probability measure on $\mathbb{X}$ with full support. Let $Q$ be the forward $\mathscr{I}_{\alpha}$-projection of $R$ on $\mathbb{L}$. Write $Q_0=R$ and write $Q_n$ for the forward $\mathscr{I}_{\alpha}$-projection of $Q_{n-1}$ on $\mathbb{L}_{n-1}$, where for $n>m$, $\mathbb{L}_n = \mathbb{L}_i$, $i = n~(\text{mod } m)$. Then $Q_n\to Q$.
\end{theorem}

\vspace*{.1in}

\begin{IEEEproof} The proof largely follows Csisz\'ar's proof of \cite[Th.~3.2]{1975xxAP_Csi} with the main changes being the use of the generalization of Pinsker's inequality \cite[Th.~1]{1967xxSSMH2_Csi} and some care to address convergence of the escort measures. Details follow.

First let us observe that if $\text{Supp}(\mathbb{L}_n) \nsubseteq \text{Supp}(Q_{n-1})$, in order to find the projection of $Q_{n-1}$ on $\mathbb{L}_n$, one may restrict attention to members $P\in \mathbb{L}_n$ with $\text{Supp}(P)\subset \text{Supp}(Q_{n-1})$. If not, $\mathscr{I}_{\alpha}(P,Q_{n-1}) = \infty$. With this restricted $\mathbb{L}_n$, by Proposition \ref{p1:prop:alg_inner_point}, $Q_n$ is an algebraic inner point of the restricted $\mathbb{L}_n$. Henceforth we call these simply $\mathbb{L}_n$ and denote their intersection by $\mathbb{L}$.

Fix a natural number $N$. In view of Proposition \ref{p1:prop:alg_inner_point}, applying Theorem \ref{p1:thm:pyth_and_proj}-(a), we see that for any $P\in \mathbb{L}$ we have
\begin{eqnarray}
\label{p1:eqn:pyth-betwenn-Q_n}
\mathscr{I}_{\alpha}(P,Q_{n-1}) = \mathscr{I}_{\alpha}(P,Q_n)+\mathscr{I}_{\alpha}(Q_n,Q_{n-1}), \quad n=1,\dots, N.
\end{eqnarray}
Summing all the $N$ equations, we get
\begin{eqnarray*}
\mathscr{I}_{\alpha}(P,R) = \mathscr{I}_{\alpha}(P,Q_N)+\sum_{n=1}^N \mathscr{I}_{\alpha}(Q_n,Q_{n-1}) \quad \forall P\in \mathbb{L}.
\end{eqnarray*}
Now let $(Q_{N_k})$ be a subsequence of $(Q_n)$ converging to, say, $\tilde{Q}$. Taking limit as $k \rightarrow \infty$ along this subsequence, we get
\begin{eqnarray}
\label{p1:equality_sum}
\mathscr{I}_{\alpha}(P,R) = \mathscr{I}_{\alpha}(P,\tilde{Q})+\sum_{n=1}^{\infty} \mathscr{I}_{\alpha}(Q_n,Q_{n-1}) \quad \forall P\in \mathbb{L},
\end{eqnarray}
which implies that the summation term is finite, and so $\mathscr{I}_{\alpha}(Q_n,Q_{n-1})\to 0$, or $I_f(Q_n',Q_{n-1}') \to 0$, as $n \rightarrow \infty$ in view of (\ref{p1:eqn:alphadiv}). Hence, by \cite[Th.~1]{1967xxSSMH2_Csi}, $|Q_n'-Q_{n-1}'|_{TV}\to 0$ as $n\to \infty$. Hence all of the sequences $(Q_{N_k}')$, $(Q_{N_k+1}')$,\dots,$(Q_{N_k+m-1}')$ converge to same $\tilde{Q}'$. Now, for any $k$,  $Q_{N_k}',Q_{N_k+1}',\dots,Q_{N_k+m-1}'$, are $m$ consecutive members of the sequence $(Q_n')$, and by the periodic construction of the $Q_n$'s, each is in one of $\mathbb{L}_0',\dots,\mathbb{L}_{m-1}'$, where $\mathbb{L}_i'=\{P'\colon P\in \mathbb{L}_i\}$ with $P'$ as in (\ref{p1:eqn:escort}). Hence $\tilde{Q}'$ is in each of them which implies $\tilde{Q}'\in \mathbb{L}'$ and $\tilde{Q}\in \mathbb{L}$. Putting $P = \tilde{Q}$ in (\ref{p1:equality_sum}), we get
\begin{eqnarray*}
\mathscr{I}_{\alpha}(\tilde{Q},R) = \sum_{n=1}^{\infty} \mathscr{I}_{\alpha}(Q_n,Q_{n-1})
\end{eqnarray*}
for this subsequential limit $\tilde{Q}$. Substituting this back in (\ref{p1:equality_sum}), we see that
\begin{eqnarray*}
\mathscr{I}_{\alpha}(P,R) = \mathscr{I}_{\alpha}(P,\tilde{Q})+\mathscr{I}_{\alpha}(\tilde{Q},R) \quad \forall P \in \mathbb{L}.
\end{eqnarray*}
By Theorem \ref{p1:thm:pyth_and_proj}, $\tilde{Q}$ is the forward $\mathscr{I}_{\alpha}$-projection $Q$ of $R$ on $\mathbb{L}$. By uniqueness of the forward $\mathscr{I}_{\alpha}$-projection, every subsequential limit equals $Q$, and so $(Q_n)$ converges to $Q$.
\end{IEEEproof}

\begin{remark}
Again, the above theorem continues to hold for $\alpha>1$ under the rather restrictive assumption that each of the forward $\mathscr{I}_{\alpha}$-projections satisfies the Pythagorean property (\ref{p1:eqn:pyth-betwenn-Q_n}) with equality.
\end{remark}



\section{Concluding remarks}
\label{p1:sec:open-questions}

We end this paper with some concluding remarks.

\begin{enumerate}
  \item The forward $\mathscr{I}_{\alpha}$-projection, in general, depends on the reference measure $\mu$. The dependence on $\mu$ however disappears as $\alpha \to 1$, and in this sense $\mathscr{I}_1$-projection or $\mathscr{I}$-projection is special.

  \item Throughout this paper, motivated by constraints induced by linear statistics, we restricted $\mathbb{E}$ to be a convex set of probability measures. But it is clear that if $p$ and $q$ are two $\mu$-densities of probability measures, and both belong to $L^{\alpha}(\mu)$, then, for positive constants $c_1$ and $c_2$, we have $\mathscr{I}_{\alpha}(c_1 p, c_2 q) = \mathscr{I}_{\alpha}(p, q)$ because $\mathscr{I}_{\alpha}$ depends only on the associated escort probability densities of the arguments, and scale factors do not affect these escort densities. It would therefore be interesting to extend our theory of the forward $\mathscr{I}_{\alpha}$-projection to general convex and closed subspaces of $L^{\alpha}(\mu)$.

  \item The above remark on the insignificance of the scaling factors suggests that perhaps the theory ought to be developed from the view point of escort distributions. However, convexity of $\mathbb{E}$ which is a natural consequence of linear statistics, may be lost in the escort domain.

  \item Is there a ``generalized'' forward $\mathscr{I}_{\alpha}$-projection $Q$ for a convex $\mathbb{E}$ that is not $L^{\alpha}(\mu)$-closed? Further, if $(P_n)$ is a sequence in $\mathbb{E}$ such that $\mathscr{I}_{\alpha}(P_n,R) \to \inf_{P \in \mathbb{E}} \mathscr{I}_{\alpha}(P,R)$ as $n \to \infty$, does $P_n$ converge to this $Q$? A careful examination of the proof of Theorem \ref{p1:thm:projection} for the case when $\alpha < 1$ shows that while one can extract a {\em unique} probability measure $Q$ that satisfies
      \[
        \mathscr{I}_{\alpha}(Q,R) \leq \lim_{k \to \infty}\mathscr{I}_{\alpha}(P_{n_k},R) = \inf_{P \in \mathbb{E}} \mathscr{I}_{\alpha}(P,R)
      \]
      for any converging subsequence of densities $(p_{n_k})$ in $L^{\alpha}(\mu)$, it is not clear if $p_n \rightarrow q$, the $\mu$-density of $Q$, in $L^{\alpha}(\mu)$. However, each subsequential limit is always a scaled version of $q$. Thus $Q$ can serve as \emph{the} generalized forward $\mathscr{I}_{\alpha}$-projection. This too suggests the benefit of a theory modulo scale factors.

  \item In Section \ref{p1:sec:projection_for_linear_family}, we considered projection on linear families. Let us highlight an open question raised in that section. Is Corollary \ref{p1:cor:linear-family-iff} applicable to a family of distributions on a product space with specified marginals? While the answer is true for $\alpha = 1$ (\cite[Cor.~3.2]{1975xxAP_Csi}), we have not been able to address the general case of $\alpha >0, \alpha\neq 1$.

  \item Suppose that we have a nested sequence $\mathbb{L}_1 \supset \mathbb{L}_2 \supset \dots$ of convex sets of probability measures absolutely continuous with respect to a common $\sigma$-finite measure $\mu$ such that the respective set of densities $\mathcal{L}_n$ is closed in $L^{\alpha}(\mu)$. Let
\[
 \mathcal{L} = \bigcap_{n=1}^{\infty} \mathcal{L}_n
\]
and assume that $\mathcal{L}$ is nonempty. Questions of interest are whether the forward $\mathscr{I}_{\alpha}$-projections of a probability measure $R$ on the sets $\mathbb{L}_n$ converge to the forward $\mathscr{I}_{\alpha}$-projection on the limiting set $\mathbb{L}$ and whether the optimal values on these sets converge to that on the limiting set. Questions of this kind have been studied for entropy by Borwein and Lewis \cite{1991xxSJO_BorLew1} and for $\phi$-entropies by Teboulle and Vajda \cite{199301TIT_TebVaj}.

  \item Can one characterize the set of all regular and subspace-transitive projection rules? We therefore wish to relax the locality axiom for projection rules. This ought to include all projection rules generated by Bregman's divergences of the sum-form and additionally the projection rule generated by $\mathscr{I}_{\alpha}$.
\end{enumerate}

\appendix

\subsection{Proof of Lemma \ref{p1:lemma:properties}:}
\label{p1:sec:app-lemma1}
These properties are well-known. We provide the proofs of a) - d) for completeness. For e) we provide a reference.

a) By Jensen's inequality, $$I_{f}(P',Q')\ge 0,$$ with equality if and only if $P'=Q'$, which holds if and only if $P=Q$. Substituting this in (\ref{p1:eqn:alphadiv}), we get $\mathscr{I}_{\alpha}(P,Q) \ge \nicefrac{1}{\rho} \cdot \log(1) = 0$ for both positive and negative $\rho$, with equality if and only if $P=Q$.

\vspace{.1in}
b) Using (\ref{p1:alphadiv_expanded_general}), we get
\begin{eqnarray}
 \label{p1:eqn:Ialpha-alternate}
\mathscr{I}_{\alpha}(P,Q) & = & -\alpha\log \left(\int q^{\alpha-1}dP\right)^{\frac{1}{\alpha-1}}+\log\left(\int p^{\alpha-1}dP\right)^{\frac{1}{\alpha-1}} + (\alpha-1)\log\left(\int q^{\alpha-1}dQ\right)^{\frac{1}{\alpha-1}}.
\end{eqnarray}
By assumption, $\mathscr{I}_{\alpha_u}(P,Q)<\infty$, where $\alpha_u > 1$. From the fact that $p$ and $q$ are in $L^{\alpha_u}(\mu)$, we have that $\int p^{\alpha-1}dP = \int p^{\alpha} d\mu$ is finite and nonzero for all $\alpha \in (1,\alpha_u]$, and the same holds for $\int q^{\alpha-1}dQ$. Using these facts in (\ref{p1:eqn:Ialpha-alternate}), we conclude that $\int q^{\alpha_u-1}dP$ is finite and nonzero, and consequently so is $\int q^{\alpha-1}dP$ for all $\alpha \in (1,\alpha_u]$. We shall now apply a result \cite[Ch.~6,~Ex.~8]{1999xxRA_Fol} which states that if $g \in L^{\beta_u}(\nu)$ for some $\beta_u > 0$ and a probability measure $\nu$, then $g \in L^{\beta}(\nu)$ for $0 < \beta < \beta_u$, and
\[
  \lim_{\beta \downarrow 0} \left( \int |g|^\beta ~d\nu \right)^{1/\beta} = \exp \left\{ \int (\log |g|) ~d\nu \right\}.
\]
By setting $\beta = \alpha -1$, and by letting $\alpha \downarrow 1$, we apply the above result on each of the terms on the right-hand side of (\ref{p1:eqn:Ialpha-alternate}) and conclude that
$\int (\log q) ~dP$, $\int (\log p) ~dP$, and $\int (\log q) ~dQ$ exist, and the right-hand side of (\ref{p1:eqn:Ialpha-alternate}) goes to
\[
  -\int (\log q) ~dP +\int (\log p) ~dP + 0 = \mathscr{I}(P\|Q).
\]
A similar argument shows that when $\mathscr{I}_{\alpha_l}(P,Q) < \infty$ for some $\alpha_l <1$, we have $\lim_{\alpha \uparrow 1} \mathscr{I}_{\alpha}(P,Q) = \mathscr{I}(P\|Q)$.

c) and d) follow directly from the definitions.

\vspace{.1in}

e) This was proved by Lutwak et al. \cite[Th.~2]{200501TIT_LutYanZha} for the scalar case and by Costa et al. \cite{200307EMMCVPR_CosHerVig} for the vector case.

\subsection{Proof of Proposition \ref{p1:prop:lsc1}:}
\label{p1:sec:app-prop2}
We shall first prove the lower semicontinuity for $\alpha < 1$: if $p_n\to p$ in $L^{\alpha}(\mu)$ then
\begin{equation}
  \label{p1:eqn:lsc-desired}
  \liminf_{n\to\infty} \mathscr{I}_{\alpha}(p_n, q) \ge \mathscr{I}_{\alpha}(p,q).
\end{equation}

Fix an $\alpha<1$; this fixes a $\rho>0$. From (\ref{p1:eqn:alphadiv}), we may write
\begin{eqnarray*}
\mathscr{I}_{\alpha}(p,q) & = & \frac{1}{\rho}\log [I_f(p',q') + 1],
\end{eqnarray*}
where $f(u) = u^{1+\rho} - 1$ for $u \geq 0$.

Let $p_n\to p$ in $L^{\alpha}(\mu)$. Then $\|p_n\| \to \|p\|>0$ and since $|p_n^{\alpha}-p^{\alpha}| \le |p_n|^{\alpha}+|p|^{\alpha}$. The generalized version of the dominated convergence theorem states that (see \cite[Ch.~2,~Ex.~20]{1999xxRA_Fol} or \cite[p.139, Problem 19]{2001xxLIES_Jon}), if $\{u_n\}$ is a sequence of measurable functions on a measurable space $(\X,\mathcal{X})$ such that $u_n\to u$ $\mu$-a.e. and if $v_n, v\in L^1(\mu)$ are such that $|u_n|\le v_n$ $\mu$-a.e. and $v_n\to v$ in $L^1(\mu)$, then $u_n\to u$ in $L^1(\mu)$. By taking, $u_n = p_n^{\alpha}$, and $v_n = |p_n|^{\alpha}+|p|^{\alpha}$, the above theorem yields $p_n^{\alpha}\to p^{\alpha}$ in $L^1(\mu)$. From these, we have
\[
 (p_n/\|p_n\|)^{\alpha} \to (p/\|p\|)^{\alpha} \mbox{ in } L^1(\mu),
\]
i.e., $p_n' \to p'$ in $L^1(\mu)$, which implies $p_n' / q' \to p' / q'$ in $L^1(Q')$. (Observe that the argument thus far does not use the assumption that $\alpha < 1$ and is therefore equally applicable for an $\alpha > 1$).

Teboulle and Vajda showed in \cite[Lemma 1]{199301TIT_TebVaj} that the mapping $h \mapsto \int f(h) ~d\nu = \int h^{1+\rho} ~d\nu$ is lower semicontinuous in $L^1(\nu)$ for a probability measure $\nu$ on $(\mathbb{X}, \mathcal{X})$. Put $h_n = p_n' / q'$, $h = p' / q'$, and $\nu = Q'$. Then, we just established in the previous paragraph that $h_n \rightarrow h$ in $L^1(\nu)$. Using (\ref{p1:eqn:If-defn}) and the lower semicontinuity result of Teboulle and Vajda, we have
\begin{equation}
  \label{p1:eqn:Teboulle-Vajda}
  \liminf_{n\to\infty} I_f(p_n',q') \ge I_f(p',q') \ge 0.
\end{equation}
Since $\nicefrac{1}{\rho}\log(\cdot + 1)$ is increasing and continuous in $[0,\infty)$, using the definition in (\ref{p1:eqn:alphadiv}), (\ref{p1:eqn:Teboulle-Vajda}) implies (\ref{p1:eqn:lsc-desired}) which establishes the lower semicontinuity result for $\alpha<1$.

We now deal with the other case. Fix $\alpha>1$. Observe that the dual space of the Banach space $L^{\alpha}(\mu)$ is $L^{\alpha}(\mu)^* = L^{\frac{\alpha}{\alpha-1}}(\mu)$, and therefore $ (q/\|q\|)^{\alpha-1} \in L^{\alpha}(\mu)^*$. Consequently, the mapping defined by
\[
  T \colon L^{\alpha}(\mu) \ni h \mapsto T(h) = \int h \cdot \left( \frac{q}{\| q \|} \right)^{\alpha - 1} d\mu \in \mathbb{R}
\]
is a bounded linear functional and therefore continuous. If $p_n\to p$ in $L^{\alpha}(\mu)$, then $\|p_n\|\to \|p\|$, and therefore $p_n/\|p_n\| \to p/\|p\|$ in  $L^{\alpha}(\mu)$. By the continuity of $T$, we have
\begin{eqnarray*}
  \int \left( \frac{p_n}{\| p_n \|} \right) \left( \frac{q}{\| q \|} \right)^{\alpha - 1} d\mu
    & = & T \left( \frac{p_n}{\| p_n \|} \right) \\
    & \rightarrow & T \left( \frac{p}{\| p \|} \right), \mbox{ as } n \rightarrow \infty, \\
    & = & \int \left( \frac{p}{\| p \|} \right) \left( \frac{q}{\| q \|} \right)^{\alpha - 1} d\mu.
\end{eqnarray*}
Taking $\nicefrac{1}{\rho}\log(\cdot)$ on both sides, and using (\ref{p1:alphadiv_linear_form_general}), we see $\mathscr{I}_{\alpha}(p_n,q)\to \mathscr{I}_{\alpha}(p,q)$ where $\mathscr{I}_{\alpha}(p,q)$ may possibly be $+\infty$.

\subsection{Proof of Proposition \ref{p1:lsc2}:}
\label{p1:sec:app-prop3}
From (\ref{p1:eqn:alphadiv}), we may write
\begin{eqnarray}
\label{p1:eqn:alphadiv_intermsof_q'_p'}
\mathscr{I}_{\alpha}(p,q) = \frac{1}{\rho}\log [\text{sgn} (\rho )\cdot I_{\tilde{f}}(q',p') + 1],
\end{eqnarray}
where $\tilde{f}(u)=\text{sgn}(\rho)\cdot (u^{-\rho} - 1)$, $u \geq 0$.

Let $q_n\to q$ in $L^{\alpha}(\mu)$. Then, as in the proof of Proposition \ref{p1:prop:lsc1}, we have that $q_n'/p' \to q'/p'$ in $L^1(P')$. Following the argument of Proposition \ref{p1:prop:lsc1}, we apply the lower semicontinuity result of Teboulle and Vajda \cite[Lemma 1]{199301TIT_TebVaj} with $\tilde{f}$ playing the role of $f$, and we have
\begin{eqnarray}
\label{p1:eqn:Teboulle-Vajda-2}
 \liminf_{n\to \infty}I_{\tilde{f}}(q_n',p')\ge I_{\tilde{f}}(q',p') \geq 0.
\end{eqnarray}
If either (a) $\rho < 0$ and $I_{\tilde{f}}(q', p') = 1$, or (b) $\rho > 0$ and $I_{\tilde{f}}(q', p') = \infty$, then using the first inequality in (\ref{p1:eqn:Teboulle-Vajda-2}) and using (\ref{p1:eqn:alphadiv_intermsof_q'_p'}) one easily verifies the limit
\begin{eqnarray}
  \label{p1:eqn:lsc-q-1}
  \liminf_{n\to \infty}\mathscr{I}_{\alpha}(p,q_n) = \mathscr{I}_{\alpha}(p,q) = +\infty.
\end{eqnarray}
For all other cases, we recognize that $\nicefrac{1}{\rho} \log [\text{sgn}(\rho)\cdot u + 1]$ is an increasing continuous function for $u \in [0,1]$ when $\rho<0$ and for $u \in [0,\infty)$ when $\rho>0$. Using this, the first inequality in (\ref{p1:eqn:Teboulle-Vajda-2}), and (\ref{p1:eqn:alphadiv}), we have the following analog of (\ref{p1:eqn:lsc-desired})
\begin{eqnarray}
  \label{p1:eqn:lsc-q-2}
  \liminf_{n\to \infty}\mathscr{I}_{\alpha}(p,q_n) \ge \mathscr{I}_{\alpha}(p,q).
\end{eqnarray}
Equations (\ref{p1:eqn:lsc-q-1}) and (\ref{p1:eqn:lsc-q-2}) together establish the lower semicontinuity in the second argument.

\subsection{Proof of Proposition \ref{p1:quasiconvexity}:}
\label{p1:sec:app-prop4}
Let $p_0,p_1\in \overline{B}(q,\tau)$, i.e., using (\ref{p1:alphadiv_linear_form_general}),
\begin{equation}
  \label{p1:eqn:within-ball}
  \text{sgn}(\rho)\int \frac{p_\lambda}{\|p_\lambda\|} \left( \frac{q}{\|q\|} \right)^{\alpha-1} d\mu \le \text{sgn}(\rho)\cdot t \quad \text{for } \lambda=0,1,
\end{equation}
where $t=\exp\{\tau \rho\}$. Now, let us consider $\lambda \in [0,1]$, and define
\begin{equation}
  \label{p1:eqn:p-lambda}
  p_{\lambda} := \lambda p_1 + (1-\lambda) p_0.
\end{equation}
We then have the following chain of inequalities:
\begin{eqnarray*}
 \lefteqn{\text{sgn}(\rho)\int \frac{p_{\lambda}}{\|p_{\lambda}\|} \left( \frac{q}{\|q\|} \right)^{\alpha-1} d\mu}\\
& \stackrel{(a)}{=} & \frac{\text{sgn}(\rho)}{\|p_{\lambda}\|}\left[\lambda \int p_1\left( \frac{q}{\|q\|} \right)^{\alpha-1} d\mu + (1-\lambda)\int p_0\left( \frac{q}{\|q\|} \right)^{\alpha-1} d\mu\right]\\
& \stackrel{(b)}{\le} & \frac{\text{sgn}(\rho)}{\|p_{\lambda}\|} \left[\lambda \|p_1\| t +(1-\lambda)\|p_0\| t \right]\\
& = & \text{sgn}(\rho)\cdot t \cdot \frac{\left[\lambda \|p_1\| +(1-\lambda)\|p_0\| \right]}{\|p_{\lambda}\|} \\
& \stackrel{(c)}{\le} & \text{sgn}(\rho)\cdot t \cdot 1,
\end{eqnarray*}
where (a) follows by plugging in (\ref{p1:eqn:p-lambda}), (b) follows from (\ref{p1:eqn:within-ball}), and (c) follows because Minkowski's inequality gives that, for $\alpha > 1$, $\|p_{\lambda}\|\le \lambda \|p_1\| +(1-\lambda)\|p_0\|$ while for $0 < \alpha < 1$, this inequality is reversed.

Using (\ref{p1:alphadiv_linear_form_general}) once again, this time to write the above inequality in terms of $\mathscr{I}_{\alpha}$, we get $\mathscr{I}_{\alpha}(p_\lambda,q)\le \tau$, which implies $p_\lambda \in \overline{B}(q,\tau)$ for $\lambda \in [0, 1]$.

\subsection{Counterexamples as indicated Remark \ref{p1:remark:countereg}:}
\label{p1:sec:app-countereg}

Let $\X = (0,1)$. Let $\mu = Q$ be the Lebesgue measure on $\X$. Let
\begin{equation}
\mathbb{L} = \left\{P\colon\int f_n\, dP = 0, n = 1,2,3,\dots\right\},
\end{equation}
where
\begin{eqnarray}
 \label{p1:eqn:f-n}
  f_n(x) := \left\{ \begin{array}{ll}
    \frac{3+\sqrt{n}}{4} & 0 < x < \frac{1}{4n} \vspace{.05in}\\
    \frac{3}{4} & \frac{1}{4n} \le x < \frac{1}{4} \vspace{.05in}\\
    -\frac{1}{4} - \frac{1}{4\sqrt{n}} & \frac{1}{4} \le x < \frac{1}{2} \vspace{.05in}\\
    -\frac{1}{4} & \frac{1}{2} \le x < 1.
  \end{array}
  \right.
 \end{eqnarray}
Then $Q\in \mathbb{L}$. Clearly $\lim_{n\to\infty}f_n = \frac{3}{4}\cdot 1_{\{(0,\frac{1}{4})\}} - \frac{1}{4}\cdot 1_{\{(\frac{1}{4},1)\}}$. Now $g = \lim_{n\to\infty}f_n$ is in the closure of the linear span of $\{f_n\}_{n\ge 1}$, but not in the linear span of $\{f_n\}_{n\ge 1}$. Let $R$ be a probability measure whose $\mu$-density $r$ satisfies $q^{\alpha -1} = c\, r^{\alpha -1} + g$. Then $c = {\int q^{\alpha} d\mu}/{\int q r^{\alpha -1}}$. Notice that the inequality in (\ref{p1:eqn:pythagorean_inequality}), using (\ref{p1:alphadiv_linear_form_general}), is equivalent to 
\begin{eqnarray}
\label{p1:eqn:pythagorean_inequality-app1}
\int p (q^{\alpha -1} - cr^{\alpha -1}) d\mu & \ge & 0 \text{ if } \alpha >1\\
\label{p1:eqn:pythagorean_inequality-app2}
& \le & 0 \text{ if } \alpha <1.
\end{eqnarray}
$\alpha >1$: {\em Necessary condition is not sufficient}: Let $P$ be a probability measure defined by
\begin{eqnarray}
    \label{p1:eqn:dp/dq}
     \frac{dP}{dQ}(x) =  \left\{ \begin{array}{rll}
     \frac{1}{5\sqrt{x}} & 0 < x < \frac{1}{4}\vspace{.05in} \\
     0 & \frac{1}{4} \le x < \frac{3}{5}\vspace{.05in} \\
     2 & \mbox{ if } \frac{3}{5} \le x < 1.
    \end{array}
\right.
  \end{eqnarray}
It is easy to check that $P\in \mathbb{L}$. The left-hand side of (\ref{p1:eqn:pythagorean_inequality-app1}), for the $P$ defined above, evaluates to $-\nicefrac{1}{20} \ngeq 0$. Therefore, by Th. \ref{p1:thm:pyth_and_proj}, $Q$ cannot be the forward $\I_{\alpha}$-projection of $R$ on $\mathbb{L}$.

\vspace{0.2cm}

$\alpha >1$: {\em Sufficient condition is not necessary}: Define $R$ by setting $g = -\lim_{n\to\infty}f_n$. The left-hand side of (\ref{p1:eqn:pythagorean_inequality-app1}) is
\begin{eqnarray*}
\int g\, dP & = & -\int (\lim_{n\to \infty}f_n)\, dP\\
 & \ge & - \lim_{n\to \infty}\int f_n \, dP\\
 & = & 0,
\end{eqnarray*}
where the last inequality follows by Fatou's lemma. Since this holds for every $P\in \mathbb{L}$, by Th. \ref{p1:thm:pyth_and_proj}, $Q$ is the $\I_{\alpha}$-projection of $R$ on $\mathbb{L}$.

\vspace{0.2cm}

For $\alpha <1$, define $R$ by setting $g = -\lim_{n\to\infty}f_n$ and $g = \lim_{n\to\infty}f_n$, respectively to show that the necessary condition is not sufficient and vice-versa.

\section*{Acknowledgements}

We thank the reviewers whose comments/suggestions helped improve this manuscript enormously.

\bibliographystyle{IEEEtran}
{
\bibliography{alpha_divergence_journal_03MAY15l}
}

\end{document}